\documentclass[preprint,showpacs,preprintnumbers,amsmath,amssymb,floatfix]{revtex4}
\usepackage{graphicx}
\usepackage{amsfonts}
\usepackage{amssymb}
\usepackage{natbib}
\usepackage{bm}
\usepackage{amsmath}

\begin{document}

\title{ Dynamical coupled-channel model of kaon-hyperon  interactions}

\vspace{0.5cm}
\author{Wen-Tai Chiang}
\affiliation{ Institute of Physics, Academia Sinica, Taipei 11529, Taiwan}

\author{B. Saghai}
\affiliation{ D\'epartement d'Astrophysique, de Physique des
Particules, de Physique Nucl\'eaire \\
et de l'Instrumentation Associ\'ee, DSM, CEA/Saclay,  
91191 Gif-sur-Yvette, France}

\author{F. Tabakin}
 \affiliation{ Department of Physics and
Astronomy, University of Pittsburgh, PA 15260, USA}

\author{T.-S. H. Lee}
 \affiliation{Physics Division, Argonne National Laboratory, Argonne,
              IL 60439, USA}

\date{\today}

\begin{abstract}
The $\pi N \rightarrow KY$ and $KY \rightarrow KY$ reactions are
studied using a dynamical coupled-channel model of meson-baryon
interactions at energies where the baryon resonances are strongly
excited. The channels included are: $\pi N$, $K\Lambda$, and
$K\Sigma$. 
The resonances considered are: 
$N^*$ [$S_{11}(1650)$, $P_{11}(1710)$, $P_{13}(1720)$,
$D_{13}(1700)$];
$\Delta^*$ [$S_{31}(1900)$, $P_{31}(1910)$, $P_{33}(1920)$];
$\Lambda ^*$ [$S_{01}(1670)$, $P_{01}(1810)$]
$\Sigma^*$ [$P_{11}(1660)$, $D_{13}(1670)$]; and $K^*$(892).
The basic non-resonant $\pi N \rightarrow KY$ and $KY\rightarrow KY$
transition potentials are derived from effective Lagrangians using
a unitary transformation method. The dynamical coupled-channel
equations are simplified by parametrizing the $\pi N \rightarrow
\pi N$ amplitudes in terms of empirical $\pi N$ partial-wave
amplitudes and a phenomenological off-shell function. Two models
have been constructed. Model A is built by fixing all
coupling constants and resonance parameters using SU(3) symmetry,
the Particle Data Group values, and results
from a constituent quark model. Model B is obtained by
allowing most of the parameters to vary around the values of model
A in fitting the data. Good fits to the available data for  
$\pi^- p \to K^\circ \Lambda,~K^\circ \Sigma^\circ$ have been achieved. 
The investigated kinematics region in the center-of-mass frame goes
from threshold to 2.5 GeV. The constructed models can be imbedded
into associated dynamical coupled-channel studies of kaon photo- and
electro-production reactions.
\end{abstract}

 \maketitle

%
\section{Introduction}
%
Investigation of kaon-nucleon and nucleon-hyperon interactions 
with hadronic probes has a long history in strangeness physics.
However, the interactions involving an additional relevant kaon-hyperon 
channel have received marginal attention, because of lack of
data. 
More recently,
strangeness reactions are also receiving considerable attention in
associated strangeness production with incident photon and
electron beams. With the advent of facilities such as JLab, ELSA, 
Spring-8, and GRAAL, copious and high precision data on
meson electromagnetic production on both  nucleon and nuclear
targets are becoming available. Measurements of the strangeness
associated production channels focus on the energy region of
$E_{\gamma}^{lab} \le $ 2.5 GeV, corresponding to the
total center-of-mass energy W$\le$2.3 GeV, which cover the
baryon resonances region. A result of our earlier
work~\cite{Chi01} on the $\gamma p \to K^+ \Lambda$ reaction showed 
that multi-step processes, due to the coupling with the $\pi N$
channel, can have as much as a $20\%$ effect on the total
cross-section. To investigate very recent strangeness production
data, it is necessary to extend that work, which was limited to
the $K \Lambda$ channel, to include all of the  $K\Sigma$
channels: $\gamma p \to K Y $,  with $K \equiv K^+ ,~K^\circ$ and
$Y \equiv \Lambda ,~\Sigma^\circ,~\Sigma^+$. Accordingly, a
dynamical coupled-channel investigation of these processes
requires realistic models  to describe  $\pi N \to \pi N,~ K Y$,
and $K Y\to  K Y$ processes. The purpose of this paper is to
report on our progress in this direction.

The importance of developing coupled-channel approaches to
meson-baryon reactions is summarized as follows:
\begin{itemize}
\item{Such an approach is required for a proper extraction of
fundamental resonance decay properties, which are ultimately to be
predicted by basic quark dynamics. In short, information about
baryon resonance properties can only be reliably extracted within
the context of an appropriate reaction theory. The importance of
this interplay between extraction of fundamental information and
the need for a consistent reaction theory has been emphasized by
Sato and Lee~\cite{Sat96} in the pion sector. Here we extend their
investigation to the kaon sector.}
\item {Impressive amounts of high quality data from
JLab~\cite{JLab}, ELSA~\cite{ELSA}, LEPS~\cite{LEPS}, and
GRAAL~\cite{Graal} offer us the opportunity to pin down the
underlying reaction mechanism and to study the role and/or
properties of intervening baryon resonances. Such an effort is a
prerequisite for any attempt to search for missing
resonances~\cite{MissR}. Combining models from a chiral
constituent quark formalism~\cite{Sagxx,Sagxy} with a
coupled-channel approach, as presented in this work, is expected
to provide reliable insights into the elementary strangeness
photo- and electro-production reactions.}
\end{itemize}
In recent years, coupled-channel effects on  meson-baryon
reactions with strangeness production have been investigated using
two approaches. Kaiser {\it et al.}~\cite{Kai97} applied a
coupled-channel approach with chiral SU(3) dynamics to investigate
pion- and photon-induced meson production near the $KY$ threshold.
Although their recent results~\cite{Car00} include p-wave
multipoles, and thus reproduce data slightly above the threshold
region, their chiral SU(3) dynamics model can not provide the
higher partial waves that are important in describing the data at
higher energies. Similar approaches have also been taken in
Refs.~\cite{Kri98,Oll00,Lut02,Hyo03}. Given the relevant W range 
mentioned above, their simplified dynamics represents of course
only a first step. Indeed, comparisons with the $\pi N \to K Y $
data clearly show that SU(3) models of Refs.~\cite{Kai97,Hyo03},
even when p-waves interactions are included~\cite{Car00}, greatly
miss fitting the differential cross-sections: theoretical
predictions produce  slopes opposite to that required by the data.

The second coupled-channel approach used in the literature to
investigate photon- and meson-induced reactions is based on using
effective Lagrangians along with a K-matrix method, developed by the Giessen Group~\cite{Feu98,Wal00,PM1,PM2}. In the
$K$-matrix approach, all intermediate states are put
on-energy-shell and hence possibly important off-shell dynamical
effects are not accounted for. The advantage of this K-matrix
approach is its numerical simplicity in handling a large number of
coupled channels. However, the extracted $N^*$ parameters may
suffer from interpretation difficulties in terms of existing
hadron models, as explicitly demonstrated in an
investigation~\cite{Sat96} of the $\Delta$ resonance.

In this paper, we present a dynamical coupled-channel model in
which the meson-baryon off-shell interactions are defined in terms
of effective Lagrangians. This off-shell approach is achieved by
directly extending existing dynamical models~\cite{Sat96} for $\pi
N$ scattering and pion photoproduction, to include $KY$ channels.
The main feature of our approach is that the strong interaction
matrix elements of $\pi N \rightarrow K Y $ and $K Y \rightarrow K
Y$ transition operators are derived from effective Lagrangians
using the unitary transformation method of Ref.~\cite{Sat96}. This
derivation marks our major differences with chiral SU(3)
coupled-channel models mentioned
above~\cite{Kai97,Car00,Kri98,Oll00,Lut02,Hyo03} since it allows one to 
include
all relevant higher partial waves and our approach is also
applicable at energies way above threshold. The dynamical content
of our approach is also clearly very different from the on-shell
$K$-matrix coupled-channel models~\cite{Feu98,Wal00,PM1,PM2}.

It is necessary to indicate more precisely, and within a more
general theoretical framework, the differences between our and
other approaches . Similar to the well-studied meson-exchange
models of $NN$ and $\pi N$ scattering, we also start with
relativistic quantum field theory. With a model Lagrangian, there
are two approaches for deriving models of meson-baryon scattering.
The most common approach~\cite{Kle74} is to find an appropriate
three-dimensional reduction of the ladder Bethe-Salpeter equation
of the considered model Lagrangian. Meson-baryon interactions are
then identified with the driving terms of the resulting
three-dimensional scattering equation; such as the
Blankenbecler-Sugar~\cite{bs} or Gross~\cite{gross} equations. A
fairly extensive study of the three-dimensional reductions for
$\pi N$ scattering is given by Hung {\it et al.}~\cite{Hung}.
Extending such reduction methods to derive coupled-channel
equations with $stable$ two-particle channels is straightforward.
In fact, the K-matrix coupled-channel equations employed in
Refs.~\cite{Feu98,Wal00} can be derived along this line if one
further neglects that the principal-value parts of the meson-baryon
propagators, which account for the off-shell dynamics.
The scattering equations used in SU(3) chiral models of
Refs.~\cite{Kai97,Car00,Kri98,Oll00,Hyo03} can also be derived
from the ladder Bethe-Salpeter equation using a procedure similar
to a three-dimensional reduction, although this simplification is
not spelled out explicitly by the authors. In Ref.~\cite{Lut02},
the Bethe-Salpeter equation is solved directly, but only for the
simplified case that the interaction kernel is of separable form
due to the use of contact interactions. The difficulties in
solving the Bethe-Salpeter equation, even with the ladder
approximation, are well documented~\cite{Keister}.

Alternatively, one can construct models of meson-baryon
interactions by deriving an effective Hamiltonian $H_{eff}$
defined in a chosen channel-space from a specific model
Lagrangian. The scattering equation within the considered
channel-space is then governed by standard scattering theory
\begin{eqnarray}
S_{\alpha\beta}(E) &=& \delta_{\alpha,\beta} - 2\pi i T_{\alpha\beta}(E) , \\
T_{\alpha\beta}(E)&=&
<\alpha\mid H_I + H_I \frac{1}{E-H_0-H_I + i\epsilon}H_I \mid \beta >
,
\end{eqnarray}
where $\alpha,\beta$ represent the relevant channels, $S$ is the
S-matrix and $T$ is the scattering operator. Here we have defined
$H_{eff}= H_0 + H_I$, with $H_0$ denoting the free Hamiltonian and
$H_I$ defining the interactions between considered channels. This
approach for $\pi N$ scattering and pion photo- and
electro-production reactions has been pursued by Sato and
Lee~\cite{Sat96}. They applied the unitary transformation method
of Refs.~\cite{Sat91,SKO} to derive $H_{eff}$ in a $\Delta\oplus
\pi N\oplus \gamma N $ channel-space. The essential idea of the
unitary transformation method we adopt is to eliminate unphysical
vertex interactions $MB \to B^\prime$ with $m_M + m_B <
m_{B^\prime}$ from the original field theory Hamiltonian (which
can be constructed from a starting model Lagrangian using the
standard canonical quantization procedure) and absorb their
effects into $MB \to M^\prime B^\prime$ two-body interactions of
the resulting $H_{eff}$. For the  $\pi N$ scattering in the
$\Delta$ region, the resulting effective Hamiltonian of the
Sato-Lee model is
\begin{eqnarray}
H_{eff} = H_0 + v_{\pi N, \pi N} + \Gamma_{\Delta \leftrightarrow \pi N} ,
\end{eqnarray}
where $v_{\pi N, \pi N}$ is the non-resonant interaction and the
$\Delta$ excitation is described by the vertex interaction
$\Gamma_{\pi N\leftrightarrow \Delta}$. With the Hamiltonian
Eq.(3), it is straightforward (as explained in Ref.~\cite{Sat96})
to show that the solution of Eq.(2) can be cast into the following
form:
\begin{eqnarray}
T_{\pi N, \pi N}(E) = t_{\pi N,\pi N}(E) + t^{R}_{\pi N,\pi N}(E)
,
\end{eqnarray}
where the non-resonant scattering operator $t_{\pi N,\pi N}$
is defined by
the
non-resonant potential $v_{\pi N, \pi N}$,
\begin{eqnarray}
t_{\pi N,\pi N}(E) = v_{\pi N,\pi N}
+ v_{\pi N,\pi N} G_{\pi N}(E)t_{\pi N,\pi N}(E) ,
\end{eqnarray}
where the $\pi N$ propagator is defined by
\begin{eqnarray}
G_{\pi N}(E) = \frac{1}{E - E_\pi({\bf k}) - E_N({\bf p}) +
i\epsilon}\  ,
\end{eqnarray}
with $E_\alpha ({\bf p}) =\sqrt{{\bf p}^2+ m_\alpha^2}$.
The resonant amplitudes (in the center-of-mass frame) is
\begin{eqnarray}
t^{R}_{\pi N,\pi N}(E) = \frac{\bar{\Gamma}^\dagger_{\Delta,\pi
N}(E)
 \bar{\Gamma}_{\Delta,\pi N}(E)}{E - m^0_\Delta -\Sigma_\Delta(E)}\ ,
\end{eqnarray}
with
\begin{eqnarray}
\bar{\Gamma}_{\Delta,\pi N}(E) &=& \Gamma_{\Delta\rightarrow\pi N}
+ \Gamma_{\Delta\rightarrow\pi N}G_{\pi N}(E) t_{\pi N,\pi N}(E), \\
\Sigma_\Delta(E) &=& \bar{\Gamma}_{\pi N}(E)G_{\pi N}(E)
\Gamma_{\pi N\rightarrow \Delta} \  .
\end{eqnarray}
It is clear from the above equations that the resonant operator  $t^R$
contains off-shell effects due to the non-resonant interaction
$v_{\pi N, \pi N}$.
Such off-shell effects must be accounted for
in order to determine from the
data the $bare$ vertex $\Gamma_{\Delta\leftrightarrow\pi N}$.
%
%
%
\begin{figure}[htb]
\includegraphics[width=0.5\linewidth]{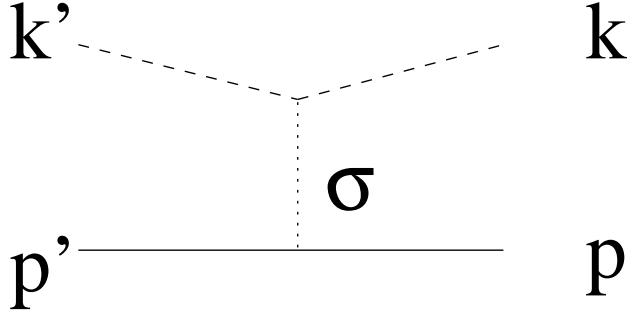}
\caption{Graphical representation for $\sigma$-exchange
$\pi N$ interaction.} \protect\label{fig:sigma}
\end{figure}
For our later discussions, we now point out that  the matrix
elements of the effective Hamiltonian, Eq.(3), can be
calculated from the usual Feynman diagrams once one specifies   the time components of the propagators of intermediate
states. For example, the $\sigma-$exchange (Fig.~\ref{fig:sigma})
of $v_{\pi N, \pi N}$ derived from the Lagrangian $L = g_{\sigma
NN} \bar{\psi}_N(x)\psi_N(x)\phi_\sigma(x)+ g_{\sigma
\pi\pi}\phi_\pi(x)\phi_\pi(x)\phi_\sigma(x)$ is of the following
form (with the normalization defined by Eqs.~(1)-(2))
\begin{eqnarray}
<{\bf p^\prime} {\bf k^\prime} \mid v^{(\sigma)}_{\pi N, \pi N}
\mid {\bf p} {\bf k} > &=&
\frac{g_{\sigma NN} g_{\sigma\pi\pi}}{(2\pi)^3}
\frac{1}{\sqrt{2E_\pi(k^\prime)}}
\sqrt{\frac{m_N}{E_N(p^\prime)}} I_\sigma
\frac{1}{\sqrt{2E_\pi(k)}}\sqrt{\frac{m_N}{E_N(p)}}
,
\end{eqnarray}
with  the propagator defined by
\begin{eqnarray}
I_\sigma =
\frac{1}{2} \left [ \frac{1}{(E_\pi(k^\prime)-E_\pi(k))^2 - {\bf q}^2
+ m_\sigma^2+i\epsilon}
+\frac{1}{(E_N(p^\prime)-E_N(p))^2 - {\bf q}^2 - m_\sigma^2 + i\epsilon}
\right ] ,
\end{eqnarray}
where ${\bf q} = {\bf k} - {\bf k}^\prime = {\bf p^\prime} - {\bf
p}$ is the three-momentum transfer. In this Hamiltonian
formulation, all particles are on their mass-shell, but the
energies are not conserved during the collisions and hence
$E_\pi(k^\prime)-E_\pi(k)\neq E_N(p^\prime)-E_N(p)$ in general.
The $\sigma$ propagator form, given in Eq.~(11), is not an
arbitrary choice, but is rigorously defined by the selected
unitary transformation. It is important to note that the matrix
element, Eq.~(10), is independent of the collision energy $E$ of
Eqs.~(1)-(2). If other methods, such as the Tamm-Dancoff method,
are chosen, the resulting effective Hamiltonian could be
energy-dependent, which then leads to non-trivial gauge invariant
problems in applying the model to study meson photo- and
electro-production reactions. The energy independence of the
resulting $H_{eff}$ is an important feature of the unitary
transformation method developed in Refs.~\cite{Sat96,Sat91}. In
this work we extend Eqs.~(3)-(9) to include $KY$ channel and higher
mass nucleon and hyperon resonances. The starting Lagrangians will
be given later. The resulting effective $v_{\pi N, KY}$ and
$v_{KY,KY}$ can be calculated from Feynman amplitudes with the
rules illustrated in Eqs.~(10)-(11).

Our goal is to construct models for describing all available data
of differential cross-sections and polarization observables of the
$\pi N \to K Y $ reactions in the total center-of-mass
energy range of W$\approx$1.3 to 2.3 GeV. 
These data~\cite{Kna75,Bak78L,Bak78S,Sax80,Har80,Datax} have been
obtained a few decades ago with low intensity beams and therefore
are not very extensive and not of high quality. Nevertheless, we
will show that they give sufficient constraints on constructing
models of $KY$ interactions.

In section II, we present the dynamical coupled-channel equations and explain
our strategy in solving these equations.
The results are given in section III. Section IV is devoted to
Summary and Conclusions.
%
%
\section{Dynamical Coupled-Channel Equations}
\protect\label{sec:Theo}

In this work, we consider a coupled-channel formulation obtained
by extending Eqs.~(3)-(9) to include the $KY$ channels.
 Specifically, we are interested in solving
\begin{eqnarray}
T_{\alpha,\beta}(E) = t_{\alpha,\beta}(E) + t^{R}_{\alpha,\beta}(E) ,
\end{eqnarray}
where $\alpha,\beta \equiv \pi N, KY$. The non-resonant scattering operator 
is defined by
\begin{eqnarray}
t_{\alpha,\beta}(E) = v_{\alpha,\beta}
+\sum_{\delta=\pi N, KY}
 v_{\alpha,\delta} G_{\delta}(E)t_{\delta,\beta}(E)
,
\end{eqnarray}
where the propagators are defined by
\begin{eqnarray}
G_{\pi N}(E) &=& \frac{1}{E - E_\pi({\bf k}) - E_N({\bf p}) + i\epsilon} ,
\\
G_{KY}(E) &=& \frac{1}{E - E_K({\bf k}) - E_Y({\bf p}) + i\epsilon} ,
\end{eqnarray}
with $E_\alpha ({\bf p}) =\sqrt{{\bf p}^2+ m_\alpha^2}$.
The resonant amplitude (in the center-of-mass frame) is
\begin{eqnarray}
t^{R}_{\alpha,\beta}(E) = \sum_{N^*_i}
\frac{\bar{\Gamma}^\dagger_{N^*_i,\alpha}(E)
 \bar{\Gamma}_{N^*_i,\beta}(E)}{E - m^0_{N^*_i} -\Sigma_{N^*_i}(E)} ,
\end{eqnarray}
with
\begin{eqnarray}
\bar{\Gamma}_{N^*_i,\alpha}(E) &=& \Gamma_{N^*_i \rightarrow \alpha}
+ \sum_{\delta=\pi N, KY}\Gamma_{N^*_i\rightarrow\delta}
G_{\delta}(E) t_{\delta,\alpha}(E) , \\
\Sigma_{N^*_i}(E) &=& \sum_{\delta=\pi N, KY}
 \bar{\Gamma}_{N^*_i\rightarrow \delta}(E)G_{\delta}(E)
\Gamma_{\delta\rightarrow N^*_i} .
\end{eqnarray}
In momentum-space, the matrix element of Eq.~(13) in the center-of-mass 
frame is
\begin{eqnarray}
  t_{\beta\alpha}(\mathbf{p}^\prime,\mathbf{p},E)
  &=&v_{\beta\alpha}(\mathbf{p}^\prime,\mathbf{p}) \nonumber \\
& &
     +\sum_{\delta} \int d{\bf p}^{\prime\prime}
         v_{\beta\delta}({\bf p}, {\bf p}^{\prime\prime})
         \frac{1}{E - E_{M_\delta}({\bf p}^{\prime\prime})
        - E_{B_\delta}({\bf p}^{\prime\prime}) + i\epsilon}
         t_{\delta\alpha}(\mathbf{p}^{\prime\prime},\mathbf{p},E),
  \label{eq:ls}
\end{eqnarray}
and the matrix element of the dressed vertex interaction Eq.~(17) is
\begin{eqnarray}
\bar{\Gamma}_{N^*_i,\alpha}({\bf p},E) &=&
\Gamma_{N^*\rightarrow \alpha}({\bf p}) \nonumber \\
& &+ \sum_{\delta}
\int d {\bf p}^\prime \Gamma_{N^*_i\rightarrow\delta}({\bf p}^\prime)
 \frac{1}{E - E_{M_\delta}({\bf p}^{\prime})
        - E_{B_\delta}({\bf p}^{\prime}) + i\epsilon}
 t_{\delta,\alpha}({\bf p}^\prime, {\bf p},E)
.
\end{eqnarray}
The integrals in the above equations extend over the
relative momentum ${\bf p}$, the off-shell dynamics is hence
included in determining the scattering amplitudes. The K-matrix
coupled-channel equation limit used by others can be obtained from
the above equations  only if one keeps the on-shell part (-$i\pi
\delta(E - E_{M_\delta}({\bf p}^{\prime\prime})-E_{B_\delta} ({\bf
p}^{\prime\prime}))$ of the propagators.

\subsection{Non-resonant amplitudes}
Our first task is to define the nonresonant potentials for solving
the coupled-channel equation (19). In the $K\Lambda$
threshold energy region, it is reasonable to derive the potentials
involving the $KY$ channel using effective Lagrangians with SU(3)
symmetry. On the other hand, it is not clear how to derive $\pi N$
potential $v_{\pi N, \pi N}$ for energies well above the $\pi N$
threshold region. We thus circumvent deriving $v_{\pi N, \pi
N}$ and instead use a phenomenological procedure to include its
effect using empirical $\pi N$ amplitudes~\cite{Arn03}. 
Accordingly, the main outcome from our calculations are scattering 
operators for $\pi N \to KY$ and $KY \to KY$ transitions, 
which are also needed for dynamical coupled-channel studies of 
$\gamma N \to KY$ reactions.

To proceed, we first derive from Eq.~(13) the following equations:
\begin{eqnarray}
t_{KY,KY}(E) &=& V_{KY,KY}(E) + V_{KY,KY}(E) G_{KY}(E)t_{KY,KY}(E) , \\
t_{KY,\pi N} &=& v_{KY,\pi N} + t_{KY,KY}(E)G_{KY}(E)v_{KY,\pi N} ,
\end{eqnarray}
where the effective $KY$ potential $V_{KY,KY}(E)$ is defined by
\begin{eqnarray}
V_{KY,KY}(E)= v_{KY,KY} + v_{KY,\pi N}[G_{\pi N}(E) + G_{\pi N}(E)
t_{\pi N,\pi N}(E) G_{\pi N}(E)]v_{\pi N, KY}
,
\end{eqnarray}
with
\begin{eqnarray}
t_{\pi N,\pi N}(E)=v_{\pi N,\pi N} +v_{\pi N,\pi N}G_{\pi N}(E)
 t_{\pi N,\pi N}(E)
.
\end{eqnarray}
We see that the operators $T_{KY,KY}$ and $T_{KY,\pi N}$ can be obtained
by solving Eqs.~(21)-(23) using the matrix elements of
$v_{KY,KY}$, $v_{KY,\pi N}$ and $t_{\pi N,\pi N}$.
We will calculate $v_{KY,KY}$, $v_{KY,\pi N}$ from effective Lagrangians.
On the other hand, we will use a phenomenological procedure to
set
\begin{eqnarray}
t_{\pi N,\pi N}({\bf p}^\prime, {\bf p}, E)=
\frac{F(p^\prime)}{F(p_0)}T^{VPI}_{\pi N,\pi N}(E) \frac{F(p)}{F(p_0)}\  ,
\end{eqnarray}
where $p_0$ is the on-shell momentum defined by $E=E_N(p_0)+E_\pi(p_0)$,
$T^{VPI}_{\pi N,\pi N}(E)$ is the empirical
$\pi N$ amplitudes taken from
the dial-in program SAID~\cite{SAID}, and we have introduced an off-shell
function
\begin{eqnarray}
F({\bf p})=\left(\frac{\Lambda_c^2}{\Lambda_c^2+{\bf p}^2}\right)^2 .
\end{eqnarray}

The matrix elements of $v_{\pi N,KY}$ and $v_{KY,KY}$ are calculated from
effective Lagrangians by using the  unitary transformation method
of Ref.~\cite{Sat96}. The effective Lagangians we consider are given in
Appendix A. The resulting potentials are the following:
\begin{alignat}{4}
  &v_{KY,\,\pi N}\ &=&\ v_{N_D}\,+&\,v_{Y_E}\,&+\,v_{K^*}\,&+\,v_{Y^*_E}
  \,,\label{eq:VpiNKY} \\[1ex]
  &v_{KY,\,KY}\ &=&\ v_{N_D}\,+&\,v_{\Xi_E}\,&+\,v_\rho\,&+\,v_{\Xi^*_E}
  \,,\label{eq:VKYKY}
\end{alignat}
where $\Xi$ is a baryon with the strangeness $S=-2$ and isospin
$I=1/2$, and $\Xi^*$ its excited states; $K^*$ indicates possible
strange vector mesons including $K^*(892)$ and $K_1(1270)$; $\rho$
here stands for all possible vector mesons ($\rho,\omega,\phi$).

\begin{figure}[t]
\includegraphics[width=0.5\linewidth]{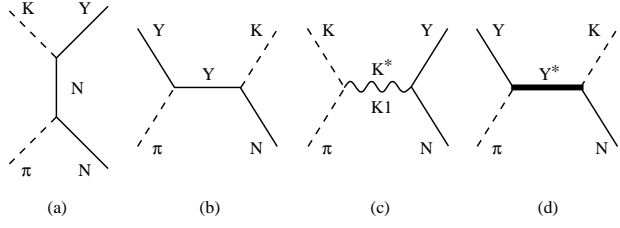}
\caption{Graphical representation of the potentials in
$\pi N \to KY$, (a) direct nucleon pole $v_{N_D}$, (b) hyperon exchange
$v_{Y_E}$, (c) strange vector meson exchange $v_{K^*}$, and (d)
hyperon resonance exchange $v_{Y^*_E}$.}
\protect\label{fig:pinky}
\end{figure}

\begin{figure}[htb]
\includegraphics[width=0.5\linewidth]{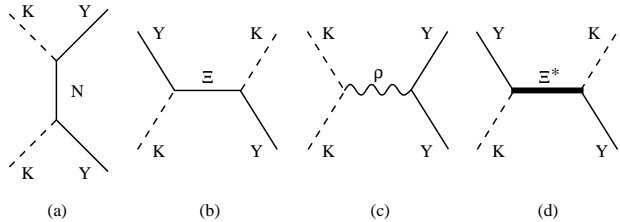}
\caption{Graphical representation of the potentials in
$KY \to KY$, (a) direct nucleon pole $v_{N_D}$, (b) $\Xi$ exchange
$v_{\Xi_E}$, (c) vector meson exchange $v_\rho$, and (d)
$\Xi$ resonance exchange $v_{\Xi^*_E}$.}
\protect\label{fig:kyky}
\end{figure}

However, not every term in Eqs.~(\ref{eq:VpiNKY}) and
(\ref{eq:VKYKY}) is computed in our calculation for a variety of
reasons. We do not consider $\Xi$ and $\Xi^*$ exchange terms,
$v_{\Xi_E}$ and $v_{\Xi^*_E}$, because of their unknown coupling
strength. The vector meson $t$-channel exchange terms, $v_\rho$
and $K_1(1270)$, are also not included because of their unknown
couplings as well as the duality hypothesis~\cite{ST96}. 
Since on the one hand, our
formalism can handle all $N^*$ resonances with spin$\leq 3/2$, 
in the $s$-channels, and on the other hand, contributions from higher 
spin $N^*$s are found~\cite{Spin5} to be negligible in the processes 
studied here, 
it should be a reasonable approximation to keep only the $t$-channel 
contributions from $K^*(892)$.
Due to the above considerations, the potentials used in this work are
\begin{alignat}{4}
  &v_{KY,\,\pi N}\ &\simeq&\ v_{N_D}\,+&\,v_{Y_E}\,&
  +\,v_{K^*(892)}\,&+\,v_{Y^*_E}
  \,,\label{eq:VpiNKY4} \\[1ex]
  &v_{KY,\,KY}\ &\simeq&\ v_{N_D}\,,\label{eq:VKYKY4'}
\end{alignat}
as illustrated in Figs.~(\ref{fig:pinky}) and~(\ref{fig:kyky}).
Their matrix elements can be calculated from
the usual Feynman
diagrams except that the propagators of intermediate particles
are defined by the procedures illustrated
in Eqs.~(10)-(11).
For $Y^*$ resonance exchange terms $v_{Y^*_E}$, the width is
included in the propagators using the
following Breit-Wigner form:
\begin{equation} 
G(\sqrt{s}) = \frac{\sqrt{\Gamma}}{\sqrt{s} - M_R +\frac{i}{2}\Gamma}\,.
\end{equation}
%
%
In appendix B, and in the next Section, we show how we determine
the coupling constants associated with the resulting potentials
using SU(3) symmetry and constituent quark models. To solve the
coupled-channel equations Eqs.~(21)-(24), the matrix elements of
the potentials must be regularized by introducing form factors,
given in Eq.~(28). The cutoff $\Lambda_c$ of these form factors
are adjusted to fit the  $\pi N \to KY$ 
data~\cite{Kna75,Bak78L,Bak78S,Sax80,Har80}.
%
%
\subsubsection{Resonant terms}
 \label{sec:Treac}
The calculation of the resonant terms $t^R_{\pi N, KY}$ and
$t^R_{KY,KY}$ using Eqs.~(16)-(18) requires bare form factors
$\Gamma_{N^*,\pi N}$ and $\Gamma_{N^*,\pi N}$ from a hadron model.
The number of the resonances we need to consider is rather large
and such calculations are not very certain at the present time. To
make progress, we postpone such that more fundamental approach and
simply adopt the following Breit-Wigner form:
\begin{equation}
t^R_{\alpha,\beta}(E)=
\sum_{N^*}\frac{\bar{\Gamma}^*_{N^*,\alpha}\,
\bar{\Gamma}_{N^*,\beta}} {E - E_{N^*}
+\frac{i}{2}\Gamma_{N^*}^{(tot)}}\,,
\end{equation}
with the total width
\begin{equation}
\Gamma_{N^*}^{(tot)} = \sum_{\alpha}
|\bar{\Gamma}_{N^*,\alpha}|^2.
\end{equation}
We will only consider the known resonances and hence the above resonant
amplitudes
can be evaluated using the information provided by the 
Particle Data
Group~\cite{PDG}. For poorly determined decay strengths 
$\bar{\Gamma}_{N^*,\pi N}$ and $\bar{\Gamma}_{N^*,KY}, $
we use a SU(3) quark model~\cite{Cap98,Kon80} to fix them.
%
%
\section{Results and discussion}
\protect\label{sec:Results and discussion}
In this Section we will first use the formalism developed in the
previous sections to build models by fitting the existing
differential cross-section and hyperon polarization asymmetry data
for the following processes:
\begin{eqnarray}
\pi^- p &\to& K^\circ \Lambda, \\
\pi^- p &\to& K^\circ \Sigma^\circ ,
\end{eqnarray}
in the center-of-mass energy region ranging from threshold to
$W \approx$ 2.3 GeV. We then present our predictions based on this
coupled-channel model for the following reactions:
\begin{eqnarray}
K^\circ \Lambda &\to& K^\circ \Lambda , \\
K^\circ \Lambda &\to& K^\circ \Sigma^\circ ,\\
K^\circ \Sigma^\circ &\to& K^\circ \Sigma^\circ .
\end{eqnarray}
To our knowledge no empirical or theoretical information about the above 
$KY \rightarrow KY$ reactions is available, although it constitutes an
important ingredient in strangeness physics, especially in dynamical 
coupled-channel studies of hyperon photoproduction reactions, as discussed
in Introduction.

To proceed, we need to construct the driving terms $v_{KY,\pi N}$, 
Figs. (2a) to (2d) and $v_{KY,KY}$, Fig. (3a), for solving 
Eqs.~(21)-(22). 

To produce numerical results for observables, the first step is to select 
a set of resonances relevant to the reaction mechanism, to be included 
in the calculation. 
To keep the number of adjustable parameters
reasonable, we need some guidance from independent investigations
on the relevant reaction mechanism, or in other words,  on the
intervening resonances in different {\it s-}, {\it u-}, and {\it
t-}channels. As mentioned in previous Sections, our final aim is to
apply  this formalism to study associated strangeness production
using electromagnetic probes. We therefore consider resonant
states that were found to be important in this realm (see e.g.
Refs.~\cite{Sagxy,SL,OS}) (though our formalism allows us to
introduce any nucleon and/or hyperon resonance with spin $\le$3/2). 
These resonances are:

\vspace{2mm}
{\it s-}channel:

$N^*$: $S_{11}(1650)$, $P_{11}(1710)$, $P_{13}(1720)$,
$D_{13}(1700)$;

$\Delta^*$: $S_{31}(1900)$, $P_{31}(1910)$, $P_{33}(1920)$.

\vspace{2mm}
{\it u-}channel:

$\Lambda^*$: $S_{01}(1670)$, $P_{01}(1810)$

$\Sigma^*$: $P_{11}(1660)$, $D_{13}(1670)$.

\vspace{2mm}
{\it t-}channel: $K^*$(892).

\noindent Notice that the above set of {\it s-}channel resonances
is in line with the findings of the Giessen Group~\cite{PM1}.

The next step is to choose the coupling constants for various
meson-baryon-baryon vertices of the mechanisms considered as shown in
Figs. (2a) to (2d) and Fig. (3a).
 
We will construct two models. 
The first model, henceforth  called model A, is obtained by fitting 
the data with most of coupling constants fixed by combining 
SU(3)-symmetry, with central values reported in the 
Particle Data Group~\cite{PDG}, and the
predictions from constituent quark models~\cite{Cap98}. 
In the second model, henceforth called model B, in fitting the data,
we let the rather 
poorly determined coupling constants used in
model A, vary within the ranges permitted by the estimated broken
SU(3)-symmetry or by the uncertainties ($\delta$) corresponding to
the ranges reported in the PDG~\cite{PDG} . 
More precisely, those adjustable parameters are allowed to vary 
within $\pm 2\delta$. Accordingly, the fixed and adjustable parameters
within our models can be classified into three sets, as explained below.

{\bf Set I:} 
The coupling constants $\pi NN$, $\pi NN^*$, and $K NY^*$
channels can be found in the literature. They are determined from
using either the SU(3) predictions or from the partial decay
widths listed by the Particle Data Group~\cite{PDG}. Those coupling
constants are listed in Table I and are used, without any adjustments, 
in constructing both models A and B.

\begin{table}[b]
\caption {Set I coupling constants taken from the SU(3)-symmetry 
predictions or PDG partial decay widths~\cite{PDG}, as discussed in 
Appendix B .}
\label{tbl:fixed}
\renewcommand{\arraystretch}{1.2}
\centering \bigskip
\begin{tabular}{ccccc}
\ Notation\ &\ Resonance\ &\  Coupling\         &\ Value\
\\ \hline
       &                   & $f_{\pi NN}$    & \ 0.997 \\[1ex]
  $N4$ &       $S_{11}(1650)~1/2^-$& $f_{\pi NN4}$ & \ 0.272 \\
  $N5$ &       $D_{13}(1700)~3/2^-$& $f_{\pi NN5}$ & \ 0.608 \\
  $N6$ &       $P_{11}(1710)~1/2^+$& $f_{\pi NN6}$ & \ 0.093 \\
  $N7$ &       $P_{13}(1720)~3/2^+$& $f_{\pi NN7}$ & \ 0.246 \\
  $L3$ & $S_{01}(1670)~1/2^-$& $f_{KN L3}$     & \ 0.078 \\
  $L5$ & $P_{01}(1810)~1/2^+$& $f_{KN L5}$     & \ 0.194 \\
  $S1$ & $P_{11}(1660)~1/2^+$& $f_{KN S1}$     & \ 0.183 \\
  $S4$ & $D_{13}(1670)~3/2^-$& $f_{KN S4}$     & \ 1.054 \\[1ex]
\end{tabular}
\end{table}

{\bf Set II:} This set includes the following coupling constants: 
$KYN$, $KYN^*$, $KY \Delta ^*$, $\pi YY$, and $\pi YY^*$.
%
\begin{table}[t]
\caption {Set II coupling constants in $\pi N \to KY$ and $KY \to KY$.
For model A, resonance pseudovector couplings are taken from
either the prediction of constituent quark models
(QM)~\cite{Cap98,Kon80} or PDG partial decay widths~\cite{PDG}, for
model B the values are extracted from our minimization procedure.} 
\label{tbl:MBcoupl}
\renewcommand{\arraystretch}{1.0}
\centering \bigskip
\begin{tabular}{ccccc}
  \ Notation\ &\ Resonance\ &\  Coupling\         &\ Model A\
  &Model B
  \\ \hline

       &                   &  $f_{K\Lambda N}$    &  -0.950
       &-0.610
  \\
       &                   &  $f_{K\Sigma N}$     & \ 0.270
       & 0.120
  \\
       &                   &$f_{\pi\Sigma\Lambda}$& \ 0.741
       & 0.010
  \\
       &                   & $f_{\pi\Sigma\Sigma}$& \ 0.710
       & 0.010
  \\
$N4$ & $S_{11}(1650)~1/2^-$ & $f_{K\Lambda N4}$
     & -0.204 & -0.254 \\
       &             &  $f_{K\Sigma N4}$    &
 0.0 & -0.200 \\
   $N5$ & $D_{13}(1700)~3/2^-$& $f_{K\Lambda N5}$    & -0.665 & -1.179
  \\
        &                     & $f_{K\Sigma N5}$     & 0.0    & -0.468
        \\
  $N6$ & $P_{11}(1710)~1/2^+$ &  $f_{K \Lambda N6}$ & 0.372 & 0.286
  \\
       &                      &  $f_{K\Sigma N6}$   & -0.162& -0.237
  \\
  $N7$ &       $P_{13}(1720)~3/2^+$&$f_{K\Lambda N7}$   &  -0.508 &
  -0.969 \\
       &                  &$f_{K\Sigma N7}$    & 0.507 & 0.461
  \\[1ex]
  $D1$ & $S_{31}(1900)~1/2^-$&$f_{K\Sigma D1}$   & 0.0 &
   -0.156   \\
  $D2$ & $P_{31}(1910)~1/2^+$&$f_{K\Sigma D2}$   & 0.0 &
    -0.200  \\
  $D3$ & $P_{33}(1920)~3/2^+$&$f_{K\Sigma D3}$   & -0.190 &
   -0.010  \\
  $L3$ & $S_{01}(1670)~1/2^-$&$f_{\pi\Sigma L3}$   &  -0.094 &
     -0.200  \\
  $L5$ & $P_{01}(1810)~1/2^+$&$f_{\pi\Sigma L5}$   & -0.111 &

  -0.010\\
  $S1$ &  $P_{11}(1660)~1/2^+$& $f_{\pi\Lambda S1}$  &  0.0 &
  -0.064 \\
       &                   & $f_{\pi\Sigma S1}$   &  -0.098 &
 -0.200 \\
 $S4$ &$D_{13}(1670)~3/2^-$& $f_{\pi\Lambda S4}$  & 0.977 & 0.252 \\
       &                   & $f_{\pi\Sigma S4}$   & 2.110 & 0.230
       \\
\end{tabular}
\end{table}
%
%
%
The coupling constants, $f_{KYN}$ and $f_{\pi YY}$,
needed for evaluating the Born terms, Figs. (2a) to (2c), are
not very well known. So we adopt the predicted {\it central values} 
using the SU(3) flavor symmetry with the well known~\cite{PiNN}
pion-nucleon coupling constant $f_{\pi NN}$ as input.
For the coupling constants associated with the decay of $N^*$,
$\Lambda^*$, and $\Sigma^*$ into $\pi Y$ or $K Y$, we use the
results of constituent quark models (QM)~\cite{Cap98,Kon80},
which have modest success in predicting baryon resonances and
their properties. Using the $N^*\rightarrow KY,\,\pi N$ and
$Y^*\rightarrow\pi Y,\,\bar{K}N$ decay amplitudes tabulated in
Refs.~\cite{Cap98,Kon80}, the resonance coupling constants, as
defined by the effective Lagrangians given in
Appendix~\ref{apdx:Lagrgn}, can be determined straightforwardly. 
These coupling constants are listed in the 4$^{th}$ column of Table II
and are used, with no adjustments, in our construction of model A. 
In model B, they are treated as adjustable parameters, 
within $\pm 2\delta$ as explained above.

{\bf Set III:} The set includes three categories, and were treated as free
parameters in constructing both models A and B, as listed in Table III.

{\it i)} The cutoff parameters
($\Lambda _{s}$, $\Lambda _{u}$, $\Lambda _{t}$, and $\Lambda
_{\pi N}$) were allowed to vary between 500 and 1200 MeV/c.

{\it ii)} The off-shell parameter for describing the
propagator of the spin 3/2 resonances, as introduced in
 Ref.~\cite{OS}. For simplicity, we assume this off-shell parameter,
$X$ in Table III, is the same for all three  spin 3/2 resonances considered. 

{\it iii)} The $K^* N Y$ coupling constants for evaluating $K^*$-exchange
mechanism illustrated in Fig. (2c).
%
%
\begin{table}[t]
\caption {Set III parameters, as extracted from minimizations for 
models A and B.}
\label{tbl:9fp}
\renewcommand{\arraystretch}{1.2}

\centering \bigskip
\begin{tabular}{cccc}
Parameter                & Symbol         &\ Model A\ & Model B

\\ \hline
 cut-offs & $\Lambda _{s}$ & 500.0 & 500.0 \\
  & $\Lambda _{u}$ & 730.1 & 1200.0 \\
  & $\Lambda _{t}$ & 1200.0 & 1199.6 \\
  & $\Lambda _{\pi N}$ & 1017.8 & 1199.9 \\[1ex]
 off-shell & $X$ & 1.178 & 1.484 \\[1ex]
 $K^* NY$ couplings & $f^V _{K^* N \Lambda}$ &  0.437 & 0.367 \\
                    & $f^T _{K^* N \Lambda}$ & -2.161 & -2.676 \\
                    & $f^V _{K^* N \Sigma}$  & -0.286 & -0.291 \\
                    & $f^T _{K^* N \Sigma}$  &  0.031 & 0.186 \\
\end{tabular}
\end{table}


\vspace{4 mm}

At this point, we wish to summarize the content of our models A and B,
and discuss briefly the extracted free parameters. In the fitting procedure,
to save computation time, we have used a data-base of about 500 points
for differential cross-sections and polarization asymmetries in the whole 
energy range of interest here. However, the resulting fits are compared
with the complete data-base, and a representative set of data are shown
in the rest of this Section.

\vspace{5 mm}
{\bf Model A:}

As described above, only parameters listed in Table III are varied in
constructing model A. All coupling constants for defining
potentials $v_{KY,\pi N}$ and $v_{KY,KY}$ are fixed, as
listed in Table I and the 4$^{th}$ column of Table II.
We note that the resulting cutoff parameters for model A, Table III, are
reasonable, while the $K^*NY$ parameters so determined  remain to be
examined theoretically. It is clear that model A can only give a
very qualitative description of the data. The model A gives a reduced
$\chi^2$ of 3.28. 

{\bf Model B:}

As mentioned above, the parameters in Table I are taken from the  literature
and are not adjusted.
The coupling constants listed in Table II for model A come from
the predictions of exact SU(3)-symmetry and/or
taking the central values of the partial decay widths listed by PDG~\cite{PDG}.
Since the SU(3)-symmetry is only an approximate symmetry, the predicted
values could have uncertainties of up to about $30\%$ . Furthermore, the ranges
specified by PDG for most of partial decay widths of resonances are
very large.
To obtain a better fit to the data and to shed
light on the relative importance of different resonances,
 model B is constructed by also varying the parameters
listed in Table II in fitting the data.
However, the ranges of these parameters are constrained by about
30 $\%$ deviation from exact SU(3) values or by $\pm 2 \delta$  
for central values taken from PDG.
The resulting parameters of model B are compared with the values of model A
in Tables~\ref{tbl:MBcoupl} and ~\ref{tbl:9fp}. It is clear that, according
to our study, the central values for the relevant parameters as reported
in literature, are not the most appropriate ones. However, the extracted
values, allowed to vary within the ranges established by other sources,
lead to a significantly reduced, improved $\chi ^2.$  It goes down by roughly
a factor of 2:  model B leads to $\chi^2$=1.77.

\vspace{5 mm}
In the following, we compare the results of models A and B with
relevant data.
%
%
%
\subsection{Differential cross section for
$\pi^- p \to K^\circ \Lambda,~ K^\circ \Sigma^\circ$ processes}

Differential cross-section at nine center-of mass total energies are 
shown in Figs. (4) and (5) for the reactions 
$\pi^- p \to K^\circ \Lambda,~ K^\circ \Sigma^\circ$,
respectively.

For the $\pi^- p \to K^\circ \Lambda$ channel, the model B (full curves)
results are comparable to those of the model A (dashed curves) up to 
$W \approx$ 1.7 GeV, and above $W \approx$ 2.0 GeV. In the intermediate 
energy range, the model B gives a better account of the data. However,
from $W \approx$ 1.8 GeV up, both models fail to reproduce the far  
backward angle data. 
%
\begin{figure}[t]
\includegraphics[width=0.8\linewidth]{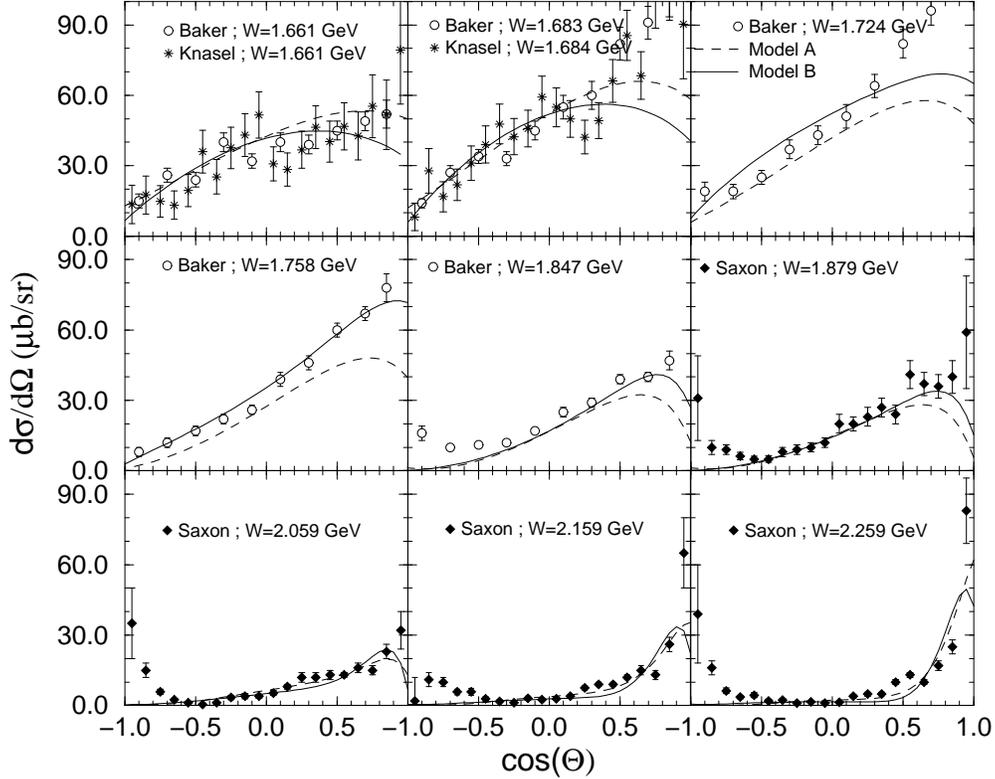}
\caption{Differential cross-section for the reaction
$\pi^- p \to K^\circ \Lambda$.
The curves are from models A (dashed curves) and B (full curves).
Data are from Refs.~\cite{Bak78L,Kna75}.}
\protect\label{fig:dsLa}
\end{figure}


For the $\pi^- p \to K^\circ \Sigma^\circ$ reaction, the situation is 
different: the model B shows a significantly better agreement with the 
data up to $W \approx$ 2.1 GeV. At the two highest energies, models 
A and B produce comparable results and they both miss the bump around
$\cos(\theta ) \approx$ 0.3.

The main gross features of our results might be explained by the ingredients
of the reaction mechanisms in our models. The $K^\circ \Lambda$ channel is
dominated by the $N^*$ resonances. In our models the included resonances 
are around M~$\approx$~1.7 GeV. To cure the above short comings, we
probably need to include higher mass resonances, especially the 
$P_{13}$(1900). This hypothesis is endorsed by the results reported in
Ref.~\cite{PM1}. In the case of the $K^\circ \Sigma^\circ$ channel, the
$\Delta$ resonances embodied in our models are around M~$\approx$~1.9 GeV
and ensure a much better reproduction of the data.

We have noticed that by loosening the constraints on the adjustable 
parameters ($\pm 3 \delta$ instead of $\pm 2 \delta$), the model-data
agreement gets improved for the $K^\circ \Lambda$ channel,
especially at backward angles. However,
we feel that the extracted values are less meaningful.

Interpretation of recent data from JLab~\cite{JLab} and ELSA~\cite{ELSA}
within a constituent quark model is in progress~\cite{Sagxy}. That work
will allow us to determine the most pertinent resonances with respect
to  strangeness electromagnetic production. Afterwards, the
present formalism will be used to imbed those resonances into planned future coupled-channel investigations of associated
strangeness photo- and electro-production. 
%
\begin{figure}[t]
\includegraphics[width=0.8\linewidth]{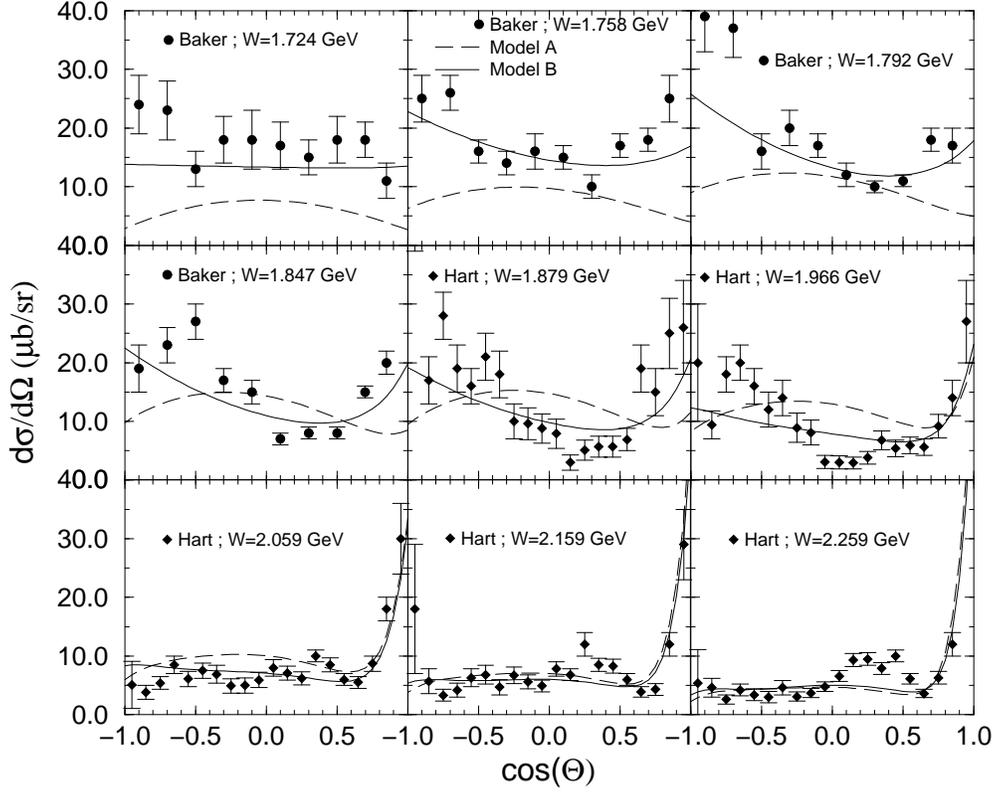}
\caption{Differential cross-section for the reaction
$\pi^- p \to K^\circ \Sigma^\circ$.
The curves are as in Fig. (4).
Data are from Refs.~\cite{Bak78S,Har80}.}
\protect\label{fig:dsS}
\end{figure}
%
\subsection{Polarization asymmetry for
$\pi^- p \to K^\circ \Lambda,~ K^\circ \Sigma^\circ$ processes}
The quality of the final state hyperon polarization asymmetry data, 
shown in Figs. 6 and 7 is clearly very poor. 
Nevertheless, as already noticed by the Giessen Group~\cite{PM1},
the inclusion of those data in the fitting procedure has a significant 
effect on the extracted coupling constants reported in Tables II and III.

%
%
\begin{figure}[t]
\includegraphics[width=0.8\linewidth]{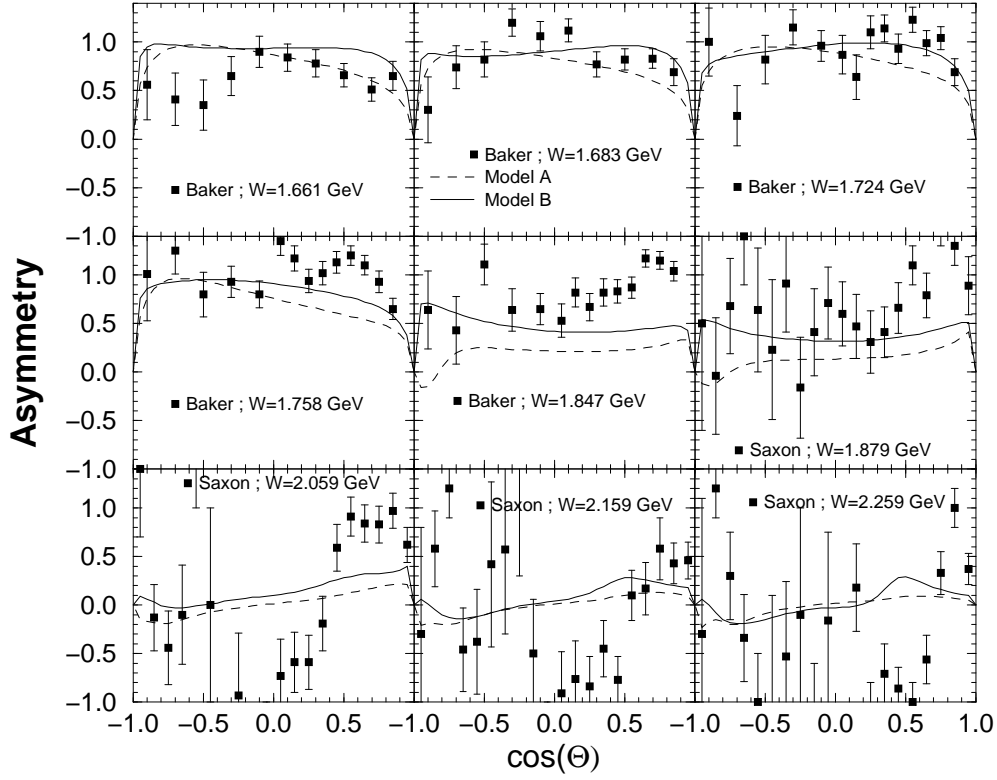}
\caption{$\Lambda$-polarization asymmetries for the reaction
$\pi^- p \to K^\circ \Lambda$.
The curves are as in Fig.~(4). Data are from 
Refs.~\cite{Bak78L,Sax80}.}
\protect\label{fig:asyLa}
\end{figure}

The main features of the polarized $\Lambda$ asymmetries (Fig. (6))
are that they are large and positive up to $W \approx$ 1.8 GeV, and
above they show nodal structures.  The model B shows a better 
agreement with the data at lower energies. The short comings at
higher energies could again be attributed to the lack of higher mass
$N^*$s in our models.

The most noticeable differences between models A and B are in the
shapes of the $\Sigma-$ polarization asymmetry for $W \le$~ 1.8 GeV 
(Fig. 7). The higher mass $\Delta$ resonances seem to play a less 
important role here than in the case of the differential cross-sections
(Fig. 5).
%
%
\begin{figure}[t]
\includegraphics[width=0.8\linewidth]{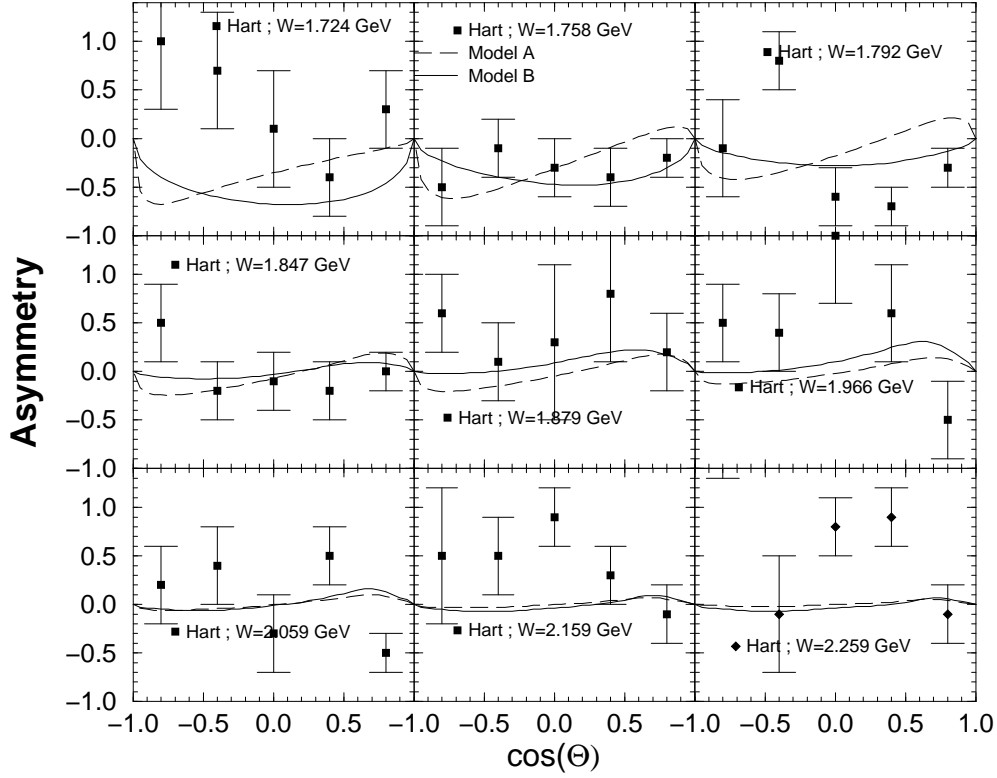}
\caption{$\Sigma$-polarization asymmetries for the reaction
$\pi^- p \to K^\circ \Sigma^\circ$.
The curves are as in Fig.~(4).
Data are from Ref.~\cite{Har80}.}
\protect\label{fig:asyS}
\end{figure}
\newpage
%
%
\subsection{Role of the nucleon resonances in  the reactions
$\pi^- p \to K^\circ \Lambda, K^\circ \Sigma^\circ$.}

It is interesting to identify the role of nucleon resonances within our
model B. To do so, we have turned off the nucleon resonances, one at
a time, by putting the relevant couplings in Table II to zero,
and have calculated the observable without any readjustment of the 
other parameters. The excitation functions, at three angles, for the 
cross-sections and the polarization asymmetries are depicted in Figures 
(8) and (9) for the reactions 
$\pi^- p \rightarrow K^\circ \Lambda\ {\rm and}\ K^\circ \Sigma^\circ$, 
respectively.

The notation used in the figures for the resonances are those in 
Table II; namely, N4: $S_{11}(1650)$, N5: $D_{13}(1700)$, 
N6: $P_{11}(1710)$, and N7: $P_{13}(1720)$.

The most striking feature here is the angular dependence of the
role played by each resonance.

In the $K^\circ \Lambda$ channel (Fig. 8, left column), 
the effects on the differential
cross-sections due to the $S_{11}(1650)$
goes from highly dominant at forward angles to marginal at large 
backward angles. 

For the P-wave resonances, $P_{11}(1710)$ and $P_{13}(1720)$,
we observe strong effects at extreme angles, which also reveal  
large interference phenomena.
 
The D-Wave resonance, $D_{13}(1700)$, has a significant effect only 
below $\approx$ 1.7 GeV and show a sharp increase at intermediate and 
large angles. The possible role played by D-wave resonances has not
been investigated in other recent works~\cite{Car00,PM1}.
%
\begin{figure}[b]
\includegraphics[width=0.8\linewidth]{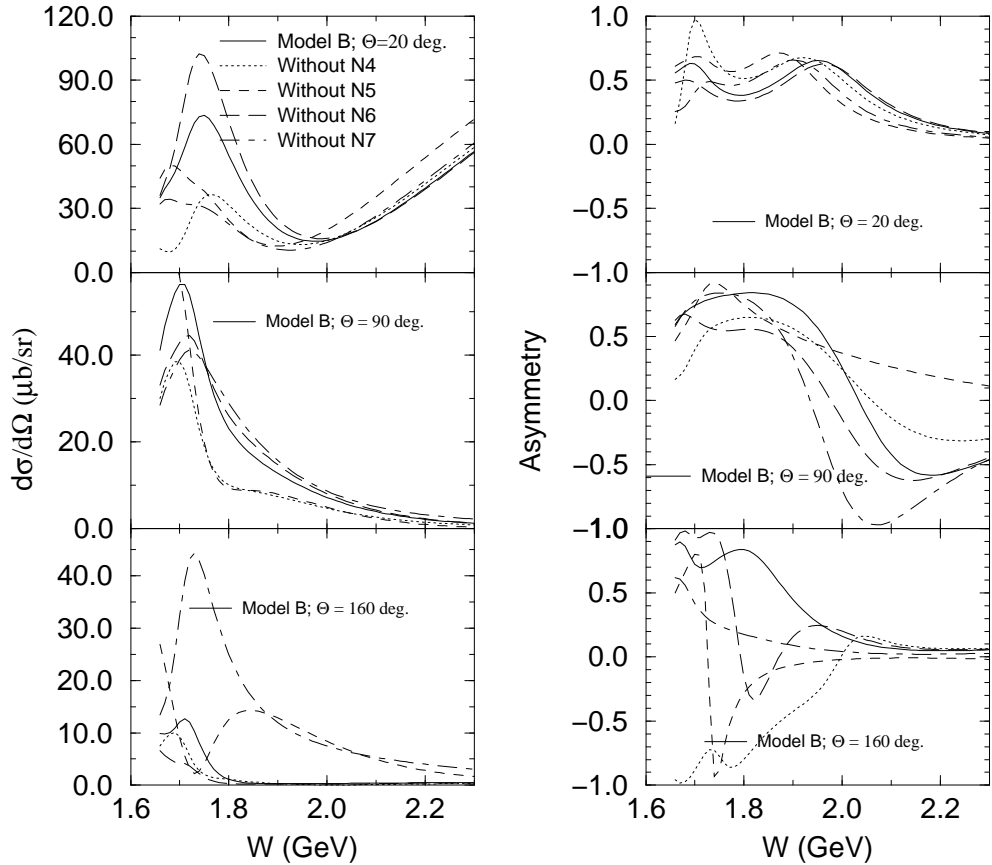}
\caption{Excitation function at three angles for the reaction
$\pi^- p \to K^\circ \Lambda$ from threshold up to W=2.3 GeV.
The curves are from model B (full curve), and the same model without
one nucleon resonance: N4 (dotted), N5 (dashed), N6 (long dashed),
N7 (dash-dotted).}
\protect\label{fig:exfL}
\end{figure}
%
%
The polarized $\Lambda$ asymmetries (Fig. 8, right column), show very
different behavior. At the forward angles,  different resonances have
comparable effects. Around 90$^\circ$, the spin 3/2 resonances, 
$D_{13}(1700)$ and $P_{13}(1720)$, produce important interferences 
effects. Finally, at large backward angles, all resonances show significant
contributions below $\approx$~1.9 GeV.

The results of a similar study on the role of the resonances for the
process $\pi^- p \to K^\circ \Sigma^\circ$ are depicted in Fig. 9.

Here the $S_{11}(1650)$ resonance has a significant effect on both
observables and at all angles. The $P_{11}(1710)$ shows small 
contributions in the whole phase space for both observables, while
the other P-wave with higher spin, $P_{13}(1720)$, produces significant
effects at forward angles in the differential cross-section
below W $\approx$ 2.0 GeV, and even more at large backward angles.
The role played by the D-Wave resonance, $D_{13}(1700)$, in the
differential cross-section increases with angle and becomes comparable
to that of $S_{11}(1650)$ at large backward angles. Finally, the polarization
observable does not show any significant sensitivity to the
$P_{13}(1720)$ and $D_{13}(1700)$ resonances.

Such a partial-wave decomposition has been also performed by the Giessen 
Group~\cite{PM1} on the total cross-section observables, leading also to
small contributions from the $P_{11}$ resonances. Effects found there
for the other three resonances are compatible with our findings.

Finally, we have performed a similar decomposition for the $\Delta$
resonances included in our model B. However, no note-worthy effect
was observed.
%
\begin{figure}[t]
\includegraphics[width=0.8\linewidth]{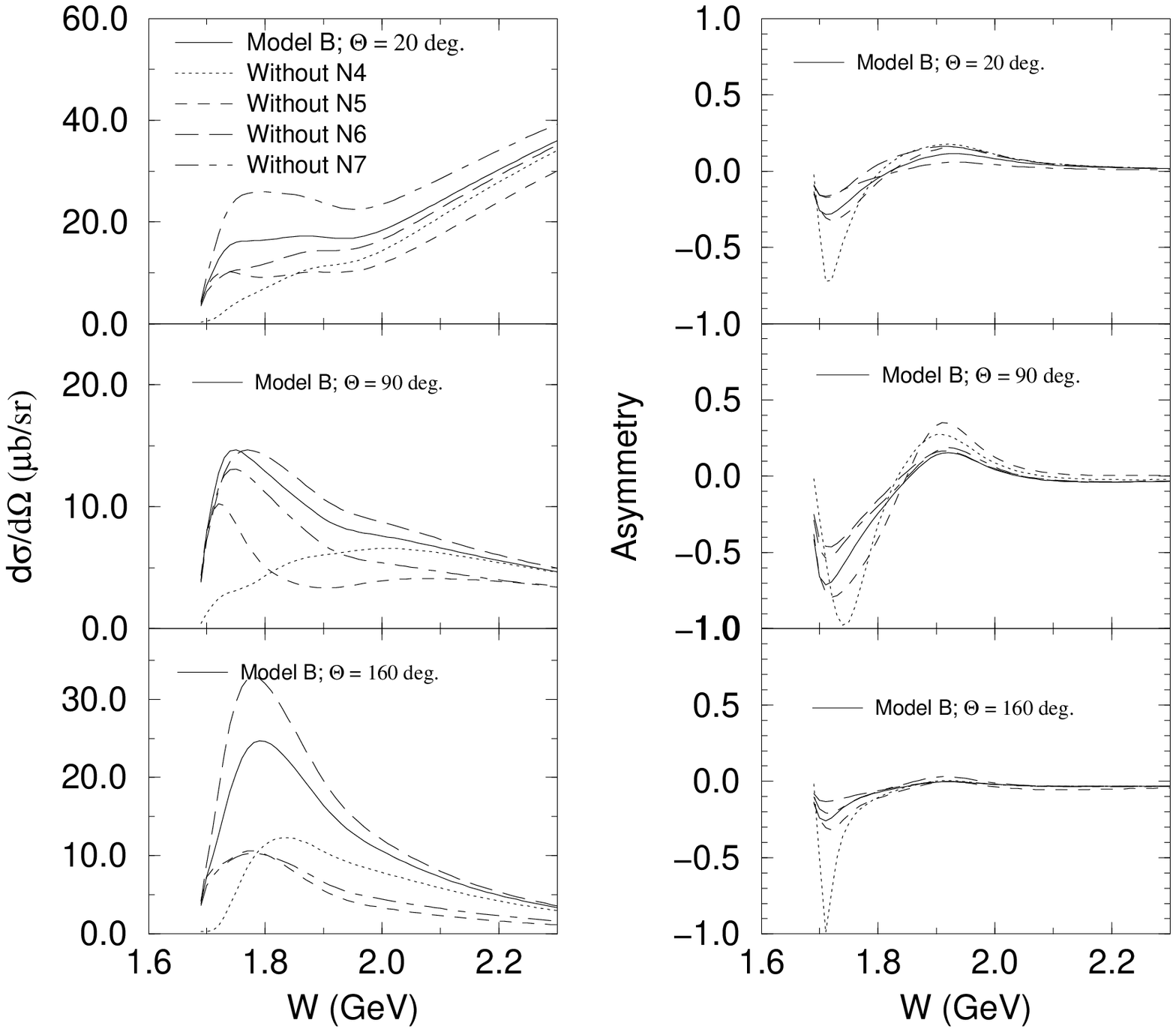}
\caption{Same as Fig. (8), but for the
$\pi^- p \to K^\circ\Sigma^\circ$ channel.}

\protect\label{fig:exfS}
\end{figure}
%
\subsection{Total cross section for the reactions
$\pi^- p \to K^\circ \Lambda$ and
$\pi^- p \to K^\circ \Sigma^\circ$}
Total cross-section data were not included in our fitting data-base.
Our results are hence, postdictions.

For the reaction $\pi^- p \to K^\circ \Lambda$ the two models give
comparable results, and  model B does slightly better at lower
energies. 

In the case of $\pi^- p \to K^\circ \Sigma^\circ$ channel the situation
is very different:  model B gives a significantly better agreement
with the data than does model A.

Both features reflect our comments about the differential cross-sections,
showing that the method used to extracted total cross-section data from 
differential cross-section measurements is sufficiently reliable. 
%
\begin{figure}[b]
\includegraphics[width=0.8\linewidth]{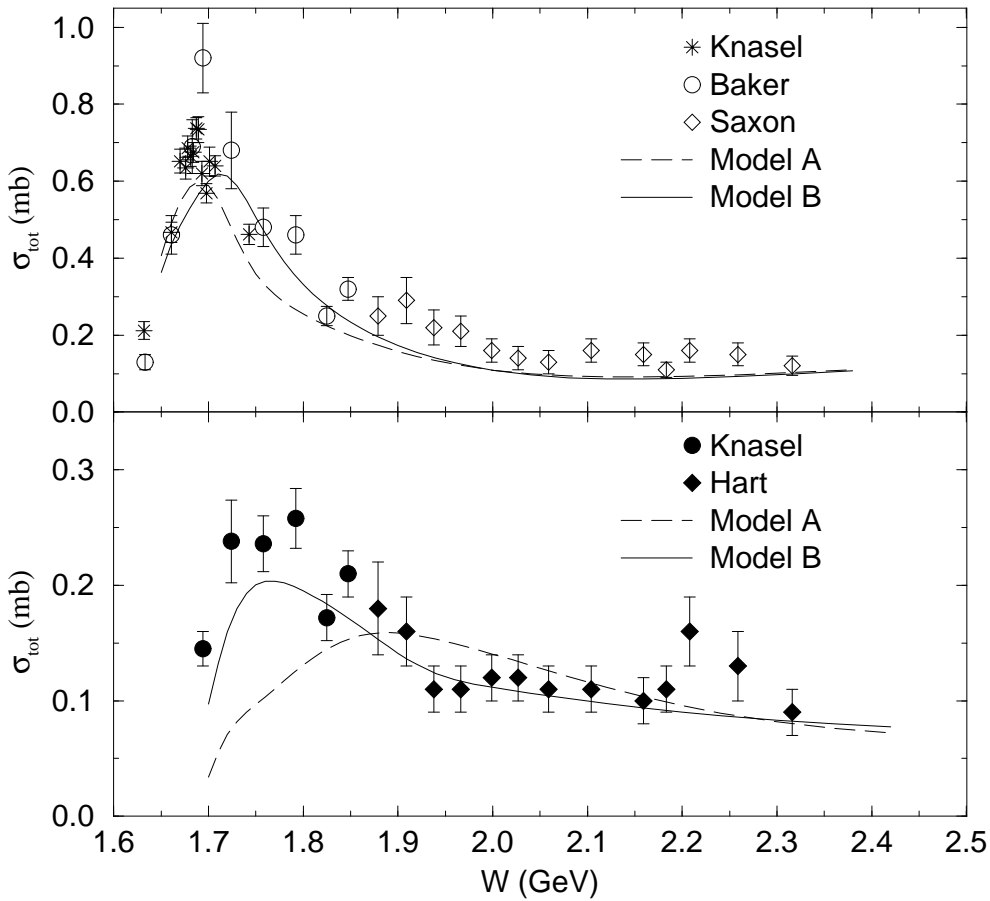}
\caption{Total cross section for the reactions
$\pi^- p \to K^\circ \Lambda$ (upper box), and
$\pi^- p \to K^\circ \Sigma^\circ$ (lower box).
Curves are the same as in Fig.~(4).
Data are from Refs.~\cite{Bak78L,Kna75,Sax80,Har80}.}
\protect\label{fig:totLS}
\end{figure}
\newpage
%
\subsection{Total cross sections of $KY \rightarrow KY$ processes}

Using the models we have constructed, one can predict the
$KY \rightarrow KY$ amplitudes.  These amplitudes,  although  presently inaccessible
experimentally,  are needed for  dynamical coupled-channel
investigations of the electromagnetic production of hyperons. As an example,
we show in Fig. 11 the predicted total cross sections for the
$K^\circ \Lambda \to K^\circ \Lambda$,
$K^\circ \Lambda \to K^\circ \Sigma^\circ$, and
$K^\circ \Sigma^\circ \to K^\circ \Sigma^\circ$ processes.

For each of the models, we show two curves: i) contributions due only
to the resonant terms (dotted curve for model A and dash-dotted for model B),
ii) full calculation (dashed curves for model A and full curves for model B).
%
\begin{figure}[b]
\includegraphics[width=0.8\linewidth]{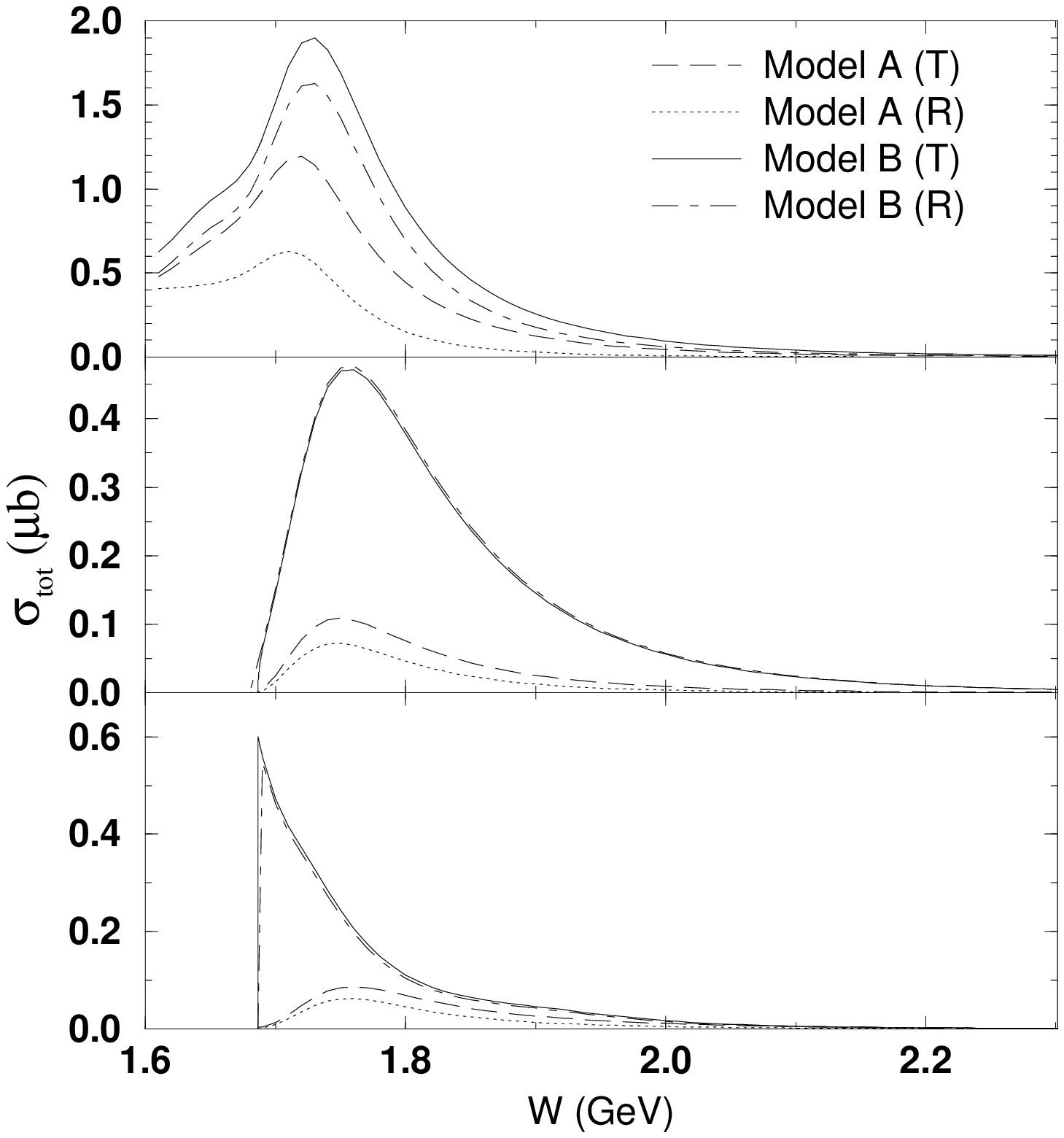}
\caption{Total cross section for the reactions
$K^\circ \Lambda \to K^\circ \Lambda$ (upper box),
$K^\circ \Lambda \to K^\circ \Sigma^\circ$ (middle box),
and
$K^\circ \Sigma^\circ \to K^\circ \Sigma^\circ$ (lower box).
Curves come from only resonant terms for models A and B (dotted and
dash-dotted, respectively, and full A and B models (dashed and
full curves, respectively).}

\protect\label{fig:totKY}
\end{figure}
%

We see that the predictions from model A and model B are strikingly 
different. Within model B, the resonant terms play a more significant 
role in all three channels. Moreover, the magnitude of the total
cross sections are higher by roughly a factor of 4 for model B than
for model A.

We therefore expect that the more realistic model B will generate very 
different final state $KY$ scattering effects on Kaon electromagnetic 
production reactions.
Our investigations in this direction will be
published elsewhere.
\section{Summary and Conclusions}
Based on an extension of the dynamical model of Ref.~\cite{Sat96},
we have developed an approach to construct the coupled-channel models for
describing the $\pi N \rightarrow KY$ and $KY \rightarrow KY$ reactions 
at energies where the baryon resonances are strongly excited. 
As a start, we only consider
$\pi N$ and $KY$ ($\equiv~K\Lambda$, $K\Sigma$) channels. Furthermore, 
the resonances which were found to be important in the $\pi N \to K Y$
and kaon photoproduction reactions are included in the investigations. 
Thus the  models we construct can be consistently used to
also investigate kaon electromagnetic production reactions. 
Undoubtedly, our objective is very limited compared to a more
rigorous coupled-channel approach,  which necessarily includes more
channels, such as $\pi \Delta$, $\rho N$, and $\omega N$.
However, our approach can be used to include additional nucleon and
hyperon resonances with spin $\le$ 3/2.
 
Given that no attempt is made to also fit the $\pi N$ elastic scattering 
data, we solve the coupled-channel equations with a  simplification 
that the $\pi N \rightarrow \pi N$ scattering t-matrix 
elements are parameterized in terms of the empirical $\pi N$ partial-wave
amplitudes and a phenomenological off-shell function. 
On the other hand, the basic non-resonant $\pi N \rightarrow KY$ 
and $KY\rightarrow KY$ transition potentials are derived rigorously 
from effective Lagrangians using a unitary transformation method. 

We have constructed two models. The first one (model A) is built by 
assuming that all coupling constants and resonance parameters can be 
fixed using SU(3)-symmetry information from the Particle Data Group, 
plus values from a constituent quark model. The second model (B) is 
obtained by allowing most of the parameters to vary around the values of 
model A in fitting the data. Good fits to the available differential
cross section and spin observable data for  $\pi^- p \to K^\circ
\Lambda,~K^\circ \Sigma^\circ$ have been achieved. The
investigated kinematics region in the center-of-mass frame goes
from threshold to 2.5 GeV. 

The constructed models will facilitate coupled-channel studies of 
kaon photo- and electro-production reactions. 
In particular, the predicted $KY \rightarrow KY$ amplitudes, which 
are inaccessible experimentally, are needed to predict  
coupled-channel effects,  such as that due to the 
$\gamma N \rightarrow K \Lambda \rightarrow K \Sigma$ transition.
Our effort in this direction will be published elsewhere.

\begin{acknowledgments}
                                                                                
One of us (B.S.) wishes to express appreciation for warm hospitality
during a visit to the University of Pittsburgh.
This work was supported in part by the U.S. Department of Energy,
Office of  Nuclear Physics, under Contract No. W-31-109-ENG-38 and
in part by the U.S. National Science Foundation, under Grant No.
0244526 at the University of Pittsburgh.  The gracious hospitality
during visits to Saclay and to Argonne is very much appreciated by F.T.

\end{acknowledgments}
\newpage
%
%
\begin{appendix}

%

\section{LAGRANGIANS}  \label{apdx:Lagrgn}
The effective Lagrangians used in this work are given in this
appendix for reference.

\subsection{Born term interaction} The
$0^-$meson-$\frac{1}{2}^+$baryon interactions are usually
described using either pseudoscalar(PS) or pseudovector (PV) 
coupling,
\begin{eqnarray} \label{eq:PSMBB'}
  {\cal L}_{MBB'}^{(PS)} &=& -ig_{MBB'}\,
  \bar{\psi}\,\gamma_5\,\psi'\,\phi\, + \,h.c.\,,
  \\[1em] \label{eq:PVMBB'}
  {\cal L}_{MBB'}^{(PV)} &=& -\frac{f_{MBB'}}{m_\pi}\,
  \bar{\psi}\,\gamma_5\,\gamma_\mu\,\psi'\,\partial^\mu\phi\, + \,h.c.\,.
\end{eqnarray}
If baryons $B$ and $B'$ are on-shell, then ${\cal L}_{MBB'}^{(PS)}$
and ${\cal L}_{MBB'}^{(PV)}$ are equivalent, and the pseudoscalar coupling
$g_{MBB'}$ and pseudovector coupling $f_{MBB'}$ are related by
\begin{equation} \label{eq:Bpsvs}
  \frac{f_{MBB'}}{m_\pi} = \frac{g_{MBB'}}{M_B+M_{B'}}\,.
\end{equation}
In this work, the pseudovector coupling is used for both $\pi$ and $K$
sectors. Using SU(3) symmetry as discussed later, we can express the interaction
Lagrangians in each particle basis.  For example, the Lagrangian for the
($K^+p\,\Lambda$) vertex can be written as
\begin{equation}
  {\cal L}_{K^+p\,\Lambda}^{(PV)}= -\frac{f_{K\Lambda N}}{m_\pi}
     \Bigl(\,\bar{p}\,\gamma_5\gamma_\mu\,\Lambda\,\partial^{\mu}K^+
     +\bar{\Lambda}\,\gamma_5\gamma_{\mu}\,p\,\partial^\mu\bar{K}^-\Bigr) \ ,
\nonumber
\end{equation} where the field operators are denoted by the particle's identity.

\subsection{SU(3) symmetry}
The notation used to described the particle fields is defined here:
\begin{equation}
 N \equiv \left( \begin{array}{c} p\\ n \end{array} \right) , \qquad
            \bar{N} \equiv \bigl(\bar{p},\;\bar{n}\bigr) .
\end{equation}
\begin{equation}
 K \equiv \left( \begin{array}{c} K^+\\ K^0 \end{array} \right) , \qquad
            \bar{K} \equiv \bigl(K^-,\;\bar{K^0}\bigr) .
\end{equation}
\begin{equation}
  \bm{\tau}\cdot\bm{\pi} \equiv
  \left( \begin{array}{cc}
         \pi^0 & \sqrt{2}\pi^+ \\
         \sqrt{2}\pi^- & -\pi^0
        \end{array} \right) .
\end{equation}
\begin{equation}
 \bm{\tau}\cdot\bm{\Sigma} \equiv
  \left( \begin{array}{cc}
         \Sigma^0 & \sqrt{2}\Sigma^+ \\
         \sqrt{2}\Sigma^- & -\Sigma^0
        \end{array} \right) , \qquad
  \bar{\bm{\Sigma}}\cdot\bm{\tau} \equiv
  \left( \begin{array}{cc}
         \bar{\Sigma}^0 & \sqrt{2}\bar{\Sigma}^- \\
         \sqrt{2}\bar{\Sigma}^+ & -\bar{\Sigma}^0
         \end{array} \right) .
\end{equation}
\begin{equation}
 \Delta \equiv \left( \begin{array}{c} \Delta^{++} \\ \Delta^+ \\ \Delta^0 \\ \Delta^- \end{array} \right) , \qquad
            \bar{\Delta} \equiv \bigl( \bar{\Delta}^{++},\;\bar{\Delta}^+,\;\bar{\Delta}^0,\;\bar{\Delta}^- \bigr) .
\end{equation}
Suppressing the factors $\gamma_5\gamma_{\mu}\partial^{\mu}$ for $PV$ coupling (or $i\gamma_5$
for $PS$ coupling), the explicit interaction 
Lagrangians in the SU(3) sector for octet baryons are:
\begin{equation}
\begin{split}
{\cal L}_{\pi NN}=& -\frac{f_{\pi NN}}{m_\pi}\,
  \bar{N}\bm{\tau}N\cdot\bm{\pi}
  \\
  =& -\frac{f_{\pi NN}}{m_\pi} \Bigl[\,
  \bar{p}\,p\,\pi^0 - \bar{n}\,n\,\pi^0 + \sqrt{2}\,\bar{p}\,n\,\pi^+
  + \sqrt{2}\,\bar{n}\,p\,\pi^-
  \Bigr]\,,
  \\[1em]
{\cal L}_{\pi\Lambda\Sigma}=& -\frac{f_{\pi\Lambda\Sigma}}{m_\pi}\,
  \bigl( \bar{\Lambda}\bm{\Sigma} + \bar{\bm{\Sigma}}\Lambda \bigr)
  \cdot\bm{\pi}
  \\
  =& -\frac{f_{\pi\Lambda\Sigma}}{m_\pi} \Bigl[\,
  \bar{\Lambda}\,(\Sigma^+\pi^- + \Sigma^0\pi^0 + \Sigma^-\pi^+)
  +(\bar{\Sigma}^+\pi^+ +\bar{\Sigma}^0\pi^0 +\bar{\Sigma}^-\pi^-)\,\Lambda
  \,\Bigr]\,,
  \\[1em]
{\cal L}_{\pi\Sigma\Sigma}=&\,\ i\ \frac{f_{\pi\Sigma\Sigma}}{m_\pi}\,
  \bigl( \bar{\bm{\Sigma}}\times\bm{\Sigma} \bigr) \cdot\bm{\pi}
  \\
  =& -\frac{f_{\pi\Sigma\Sigma}}{m_\pi} \Bigl[
  (\bar{\Sigma}^+\Sigma^+ - \bar{\Sigma}^-\Sigma^-)\,\pi^0
  +(\bar{\Sigma}^0\Sigma^- - \bar{\Sigma}^+\Sigma^0)\,\pi^+
  +(\bar{\Sigma}^-\Sigma^0 - \bar{\Sigma}^0\Sigma^+)\,\pi^-
  \Bigr]\,,
  \\[1em]
{\cal L}_{K\Lambda N}=& -\frac{f_{K\Lambda N}}{m_\pi}
  \Bigl[ \,\bar{\Lambda}\,(\bar{K}N) + (\bar{N}K)\,\Lambda\, \Bigr]
  \\
  =& -\frac{f_{K\Lambda N}}{m_\pi} \Bigl[
  (\bar{p}\,K^+ +\bar{n}\,K^0)\,\Lambda + \bar{\Lambda}\,(K^-p +\bar{K}^0n)
  \Bigr]\,,
  \\[1em]
{\cal L}_{K\Sigma N}=& -\frac{f_{K\Sigma N}}{m_\pi}
  \Bigl[ \,\bar{\bm{\Sigma}}\cdot(\bar{K}\bm{\tau}N)
        +(\bar{N}\bm{\tau}K)\cdot{\bm{\Sigma}}\, \Bigr]
  \\
  =& -\frac{f_{K\Sigma N}}{m_\pi} \Bigl[\;
  \bar{\Sigma}^0(K^-p - \bar{K}^0n) + \sqrt{2}\,\bar{\Sigma}^+K^0p
  + \sqrt{2}\,\bar{\Sigma}^-K^-n \\
  & \phantom{-\frac{f_{K\Sigma N}}{m_\pi} \Bigl[\,}
  + (\bar{p}\,K^+ -\bar{n}\,K^0)\,\Sigma^0 + \sqrt{2}\,\bar{p}\,K^0\Sigma^+
  + \sqrt{2}\,\bar{n}\,K^+\Sigma^- \Bigr]\,.
\label{eq:SU(3)LgnN}
\end{split}
\end{equation}
For interactions involving the $\Delta$, which is a decuplet baryon with isospin
$3/2$ , the Lagrangians are
\begin{equation}
\begin{split}
{\cal L}_{\pi N\Delta}=& \frac{f_{\pi N\Delta}}{m_\pi}
  \Bigl[ \,\bar{\Delta}^\mu\,\bm{T}\,N + \bar{N}\,\bm{T}^\dag \Delta^\mu
  \Bigr] \cdot \partial_\mu \bm{\pi}
  \,,
  \\
  =& \frac{f_{\pi N\Delta}}{m_\pi} \Bigl[\,
  - \bar{\Delta}^{++} \pi^+ p
  + \bar{\Delta}^+ \bigr(  \sqrt{\tfrac{2}{3}}\,\pi^0 p
                         - \sqrt{\tfrac{1}{3}}\,\pi^+ n \bigr) \\
  & \phantom{\frac{f_{\pi N\Delta}}{m_\pi} \Bigl[\,}
  + \bar{\Delta}^0 \bigr(  \sqrt{\tfrac{1}{3}}\,\pi^- p
                         + \sqrt{\tfrac{2}{3}}\,\pi^0 n \bigr)\,
  + \bar{\Delta}^- \pi^- n + h.c. \,
  \Bigr] \,,
  \\[1em]
{\cal L}_{K\Sigma\Delta}=& \frac{f_{K\Sigma\Delta}}{m_\pi}
  \Bigl[ \,\bar{\Delta}^\mu\,\bm{T}\cdot{\bm{\Sigma}}\;\partial_\mu K
        + \partial_\mu\bar{K}\,\bar{\bm{\Sigma}}\cdot\bm{T}\,\Delta^\mu
  \Bigr]
  \\
  =& \frac{f_{K\Sigma\Delta}}{m_\pi} \Bigl[\,
  - \bar{\Delta}^{++} \Sigma^+ K^+
  + \bar{\Delta}^+ \bigr(  \sqrt{\tfrac{2}{3}}\,\Sigma^0 K^+
                         - \sqrt{\tfrac{1}{3}}\,\Sigma^+ K^0 \bigr) \\
  & \phantom{\frac{f_{K\Sigma\Delta}}{m_\pi} \Bigl[\,}
  + \bar{\Delta}^0 \bigr(  \sqrt{\tfrac{1}{3}}\,\Sigma^- K^+
                         + \sqrt{\tfrac{2}{3}}\,\Sigma^0 K^0 \bigr)\,
  + \bar{\Delta}^- \Sigma^- K^0 + h.c. \,
  \Bigr] \,,
\label{eq:SU(3)LgnD}
\end{split}
\end{equation}
where the four-vector indices and derivatives are suppressed in the second
lines. The couplings in Eqs.(\ref{eq:SU(3)LgnN}-\ref{eq:SU(3)LgnD}) can
be related using SU(3) symmetry~\cite{Sto97}.


\subsection{Baryon resonance interaction} \label{sec:resint}
The general interaction Lagrangians for baryon resonances (for
spin-1/2 and 3/2) are described here.  As in the Born terms, the
explicit form for each SU(3) sector can be obtained by making appropriate
substitutions in Eqs.(\ref{eq:SU(3)LgnN}-\ref{eq:SU(3)LgnD}),
\begin{equation} \label{eq:L1ps}
\begin{split}
  {\cal L}_{MBR(\frac{1}{2}^\pm)}^{(PS)} =&
  -g_{MBR}\, \bar{R}\,\Gamma\,\psi\,\phi\, +\, h.c.\,,
  \\[0.5em]
    \mbox{with}\ \Gamma\equiv& \left\{\begin{array}{cl}
      i\gamma_5 & \mbox{for $R(\frac{1}{2}^+)$}\\
      1         & \mbox{for $R(\frac{1}{2}^-)$}
      \end{array} \right. .
\end{split}
\end{equation}
\begin{equation} \label{eq:L1pv}
\begin{split}
  {\cal L}_{MBR(\frac{1}{2}^\pm)}^{(PV)} =&
    -\frac{f_{MBR}}{m_{\pi}}\, \bar{R}\,\Gamma_\mu\,\psi\,\partial^\mu\phi
    \,+\, h.c.\,,
  \\[0.5em]
    \mbox{with}\ \Gamma_\mu\equiv& \left\{\begin{array}{cl}
      \gamma_5\gamma_\mu & \mbox{for $R(\frac{1}{2}^+)$}\\
      i\gamma_\mu        & \mbox{for $R(\frac{1}{2}^-)$}
      \end{array} \right. .
\end{split}
\end{equation}
where the pseudovector couplings $f_{MBR}$ and the pseudoscalar couplings
$g_{MBR}$ for resonance $R(\frac{1}{2}^\pm)$ are related by
\begin{equation} \label{eq:Rpsvs}
  \frac{f_{MBR}}{m_\pi}=\frac{g_{MBR}}{M_R\pm M_B}\,.
\end{equation}

\begin{equation} \label{eq:L3}
\begin{split}
  {\cal L}_{MBR(\frac{3}{2}^\pm)}=& \,\frac{f_{MBR}}{m_\pi}\,
  \Bigl[\,
  \bar{R}^\mu\,\Gamma\,\Theta_{\mu\nu}(Z)\,\psi\,\partial^\nu\phi
  +\bar{\psi}\,\Gamma\,\Theta_{\nu\mu}(Z)\,R^\mu\,\partial^\nu\phi^\dag\,
  \Bigr]\,,
  \\[0.5em]
    &\mbox{with}\ \Gamma\equiv\left\{\begin{array}{cl}
      1        & \mbox{for $R(\frac{3}{2}^+)$}\\
      i\gamma_5 & \mbox{for $R(\frac{3}{2}^-)$}
      \end{array} \right. ,
  \\[0.5em]
    &\mbox{and}\ \Theta_{\mu\nu}(Z) \equiv
                 g_{\mu\nu}-(Z+\frac{1}{2})\gamma_\mu\gamma_\nu\,.
\end{split}
\end{equation}

\subsection{Vector meson interaction}

For vector meson interactions, the corresponding Lagrangians are:
\begin{equation}
\begin{split}
  {\cal L}_{K^*YN}^{(V)}=& -g^V_{K^*\Lambda N}
          \bigl(\bar{N} \gamma^{\mu}\,\Lambda\,K^*_{\mu}
          +\bar{K}^*_{\mu}\,\bar{\Lambda}\,\gamma^{\mu} N\bigr) \\
   &-g^V_{K^*\Sigma N}
    \bigl(\bar{N} \gamma^{\mu}\,\bm{\tau}\cdot\bm{\Sigma}\,K^*_{\mu}
    +\bar{K}^*_{\mu}\,\bar{\bm{\Sigma}}\cdot\bm{\tau}\,\gamma^{\mu} N
    \bigr)\,, \\[1em]
  {\cal L}_{K^*YN}^{(T)}=& -\frac{g^T_{K^*\Lambda N}}{M_{\Lambda}+M_N}
   \bigl(\bar{N} \sigma^{\mu\nu} \Lambda\,\partial_{\mu}K^*_{\nu}
   +\partial_{\mu}\bar{K}^*_{\nu}\,\bar{\Lambda}\,\sigma^{\mu\nu} N\bigr)\\
   &-\frac{g^T_{K^*\Sigma N}}{M_{\Sigma}+M_N}
     \bigl(\bar{N} \sigma^{\mu\nu} \bm{\tau}\cdot\bm{\Sigma}\,
               \partial_{\mu}K^*_{\nu}
     +\partial_{\mu}\bar{K}^*_{\nu}\,\bar{\bm{\Sigma}}\cdot\bm{\tau}\,
               \sigma^{\mu\nu} N\bigr)\,,
  \\[1em]
  {\cal L}_{K_1YN}^{(V)}=& -i\,g^V_{K_1\Lambda N}
          \bigl(\bar{N} \gamma^{\mu}\gamma_5\,\Lambda\,K^*_{\mu}
          +\bar{K}^*_{\mu}\,\bar{\Lambda}\,\gamma_5\gamma^{\mu} N\bigr) \\
   &-i\,g^V_{K_1\Sigma N} \bigl(
     \bar{N} \gamma^{\mu}\gamma_5\,\bm{\tau}\cdot\bm{\Sigma}\,K^*_{\mu}
       +\bar{K}^*_{\mu}\,\bar{\bm{\Sigma}}\cdot\bm{\tau}\,
                                      \gamma_5\gamma^{\mu} N\bigr)\,,
  \\[1em]
  {\cal L}_{K_1YN}^{(T)}=& -\frac{i\,g^T_{K_1\Lambda N}}{M_{\Lambda}+M_N}
   \bigl(\bar{N} \sigma^{\mu\nu}\gamma_5\,\Lambda\,\partial_{\mu}K^*_{\nu}
   +\partial_{\mu}\bar{K}^*_{\nu}\,\bar{\Lambda}\,
                      \gamma_5\sigma^{\mu\nu} N\bigr) \\ \nonumber
   &-\frac{i\,g^T_{K_1\Sigma N}}{M_{\Sigma}+M_N}
     \bigl(\bar{N} \sigma^{\mu\nu}\gamma_5\,\bm{\tau}\cdot\bm{\Sigma}\,
               \partial_{\mu}K^*_{\nu}
     +\partial_{\mu}\bar{K}^*_{\nu}\,\bar{\bm{\Sigma}}\cdot\bm{\tau}\,
               \gamma_5\sigma^{\mu\nu} N\bigr)\,.
\end{split}
\end{equation}
%
\section{Coupling constants} \label{hadrcoupl}
\subsection{Hadronic couplings} \label{hc}

In Sec.~\ref{sec:resint}, we give the interaction Lagrangians ${\cal
L}_{MBR}$ for spin-1/2 and 3/2 baryon resonances $R$
($=N^*,\,\Delta^*,\,Y^*$), where $B=N,\,\Delta,\,Y$ and $M=\pi,\,K$. The
coupling constants in Eqs.~(\ref{eq:L1ps}-\ref{eq:L3}) can be derived from
partial widths $\Gamma$ in the decay $R \rightarrow M B$. The derivation is
straightforward, and the formulae are given here: \\
For resonances with $J^P = 1/2^\pm$,
\begin{eqnarray} \label{eq:spin1/2}
  \Gamma_{1/2^\pm} &=& C_\mathrm{iso}\, \frac{g_{MBR}^2}{4 \pi}\,
  \frac{E_B \mp M_B}{M_R}\; q \,,
  \nonumber \\
  &=& C_\mathrm{iso}\, \frac{f_{MBR}^2}{4 \pi}
  \left( \frac{M_R \pm M_B}{m_\pi} \right)^2
  \frac{E_B \mp M_B}{M_R}\; q \,,
\end{eqnarray}
and for resonances with $J^P = 3/2^\pm$,
\begin{equation} \label{eq:spin3/2}
  \Gamma_{3/2^\pm} = C_\mathrm{iso}\, \frac{f_{MBR}^2}{12 \pi m_{\pi}^2}\,
                     \frac{E_B \pm M_B}{M_R}\; q^3 \,,
\end{equation}
where $E_B$ is the energy of the final baryon, and $q$ denotes the
three-momentum of the meson and baryon in the rest frame of the decaying
resonance. $C_\mathrm{iso}$ is the isospin factor, and $C_\mathrm{iso} = 3$
for decays $N^* \rightarrow \pi N$ and $N^* \rightarrow K \Sigma$, and
$C_\mathrm{iso} = 1$ otherwise.
\subsection{$KYN$ and $\pi YY$ couplings} \label{cc}
In this Section we summarize the situation with respect to the free 
parameters 
$f_{K\Lambda N}~,~f_{K\Sigma N}~,~f_{\pi\Sigma\Lambda}$, and 
$f_{\pi\Sigma\Sigma}$, see Table II. 

Given that in the literature pseudoscalar couplings are more commonly 
used, we would like to make clear the relation between those and the
pseudovector ones used in our work.

\subsubsection{Expressions}

Actually the issues related to the use of pseudoscalar (PS) versus 
pseudovector (PV)
couplings have been discussed by several authors (see, e.g.~\cite{PSvPV,seoul}),
but at the present time there is no strong argument to prefer one to the other.

Using de Swart convention, we have the following relations for the PS couplings:

\begin{eqnarray}
g_{K \Lambda N} & =  & - \frac{g_{\pi NN}}{\sqrt 3} ~(3-2 \alpha_D) , \\
g_{K \Sigma N} &  =  & g_{\pi NN}~ (2 \alpha_D-1) , \\
g_{\pi \Lambda \Sigma} &  = &  \frac{2}{\sqrt 3} ~\alpha_D ~g_{\pi NN} , \\
g_{\pi \Sigma \Sigma} &  = & 2 ~ (1-\alpha_D) ~g_{\pi NN} , 
\end{eqnarray}
with $\alpha_D=\frac{D}{D+F}$ the standard fraction of D- and F-couplings.

The relations between PV and PS couplings are (see e.g.~\cite{seoul,Loiseau}:
\begin{eqnarray}
f_{\pi NN} & =  & \frac{m_{\pi}}{2 M_N} ~g_{\pi NN} , \\
f_{K \Lambda N} &  =  & \frac{m_{K}}{M_N + M_\Lambda} ~g_{K \Lambda N} , \\
f_{K \Sigma N} &  =  & \frac{m_{K}}{M_N + M_\Sigma} ~g_{K \Sigma N} , \\
f_{\pi \Lambda \Sigma} &  =  & \frac{m_{\pi}}{M_\Lambda + M_\Sigma} 
~g_{\pi \Lambda \Sigma} ,  \\
f_{\pi \Sigma \Sigma} &  =  & \frac{m_{\pi}}{2 M_\Sigma} 
~g_{\pi \Sigma \Sigma} . 
\end{eqnarray}
Expressions relating PS and PV couplings for $KNN^*$ and $KNY^*$, for S- and 
P-resonances, can be found in~\cite{seoul}.

\subsubsection{Numerical considerations}

 The central values of two main KYN couplings are determined using 
Eqs.~(B3) and (B4), with
$\alpha_D=0.644$ (Ref.~\cite{Don}),
and
$g_{\pi NN}=14.11$ (Ref.~\cite{PiNN}).
For those couplings, the allowed ranges in the fitting procedure are
in line with the findings of a recent work~\cite{General} based on generalized 
Goldberger-Treiman relation combined with the Dashen-Weinstein sum rule,
which puts the following uncertainties on the 
$g_{K \Lambda N}$ and $g_{K \Sigma N}$ couplings: $\pm 36\%$ and $\pm 55\%$, 
respectively. We hence find:
$$\frac{g_{K \Lambda N}}{\sqrt{4 \pi}} = -3.934 ~~;~~ -5.351~\le 
\frac{g_{K \Lambda N}}
{\sqrt{4 \pi}}~\le~-2.518$$
$$\frac{g_{K \Sigma N}}{\sqrt{4 \pi}} = 1.146 ~~;~~ 0.516~\le 
\frac{g_{K \Sigma N}}
{\sqrt{4 \pi}}~\le~1.777 .$$

Finally, concerning the two other couplings, $\pi \Lambda \Sigma$ and
$\pi \Sigma \Sigma$, the most recent 
works that we are aware of are Refs.~\cite{Loiseau,Doi} but they do
not give identical values.


\end{appendix}

\newpage



\end{document}